\documentclass[sigconf]{acmart}

\usepackage{color}
\usepackage{tabularx}
\usepackage{xspace}
\usepackage{arydshln}
\usepackage{multirow}
\usepackage{multicol}
\usepackage{array}
\usepackage{amsmath}
\usepackage{makecell}
\usepackage{subfig}
\usepackage{balance}

\usepackage{bm}
\def\Figref#1{Figure~\ref{#1}}

\def\Secref#1{Section~\ref{#1}}
\def\eqref#1{equation~\ref{#1}}
\def\1{\bm{1}}

\def\vmu{{\bm{\mu}}}

\def\vsigma{{\bm{\sigma}}}

\def\vx{{\bm{x}}}
\def\vy{{\bm{y}}}
\def\vz{{\bm{z}}}

\DeclareMathAlphabet{\mathsfit}{\encodingdefault}{\sfdefault}{m}{sl}
\SetMathAlphabet{\mathsfit}{bold}{\encodingdefault}{\sfdefault}{bx}{n}

\newcommand{\Ls}{\mathcal{L}}

\DeclareMathOperator*{\argmax}{arg\,max}
\DeclareMathOperator*{\argmin}{arg\,min}

\usepackage{array}
\usepackage{amsmath}
\usepackage{xspace}

\newcommand{\eg}{e.\,g., }
\newcommand{\ie}{i.\,e., }
\newcommand{\wrt}{w.\,r.\,t.\, }

\usepackage{acronym}
\acrodef{RS}[RS]{recommender system}
\acrodef{RSs}[RSs]{recommender systems}
\acrodef{VAEs}[VAEs]{variational autoencoders}
\acrodef{UO}[UO]{\emph{User Orientation}}
\acrodef{DUO}[DUO]{\emph{Discounted User Orientation}}
\acrodef{RRB}[RRB]{\emph{Recommendation Representation Bias}}
\acrodef{DRRB}[DRRB]{\emph{Discounted Recommendation Representation Bias}}
\acrodef{GRU}[GRL]{Gradient reversal layer}
\acrodef{ML-1M}[ML-1M]{MovieLens-1M}
\acrodef{NDCG}[NDCG]{Normalized Discounted Cumulative Gain}
\acrodef{BA}[BA]{Balanced Accuracy}
\newcommand{\multvae}{\textsc{MultVAE}\xspace}
\newcommand{\multvaebest}{$\textsc{MultVAE}_{\textsc{Best}}$\xspace}
\newcommand{\multvaelast}{$\textsc{MultVAE}_{\textsc{Last}}$\xspace}
\newcommand{\model}{\textsc{Adv-MultVAE}\xspace}

\newcommand{\acc}{\text{Acc}\xspace}
\newcommand{\bacc}{\text{BAcc}\xspace}

\newcommand{\ndcg}{\text{NDCG}\xspace}
\newcommand{\recall}{\text{recall}\xspace}
\newcommand{\Recall}{\text{Recall}\xspace}

\newcommand{\ndcgk}{\text{NDCG}@\ensuremath{k}\xspace}
\newcommand{\recallk}{\text{recall}@\ensuremath{k}\xspace}

\newcommand{\movielens}{\ac{ML-1M}\xspace}
\newcommand{\lfm}{\textsc{LFM-2b}\xspace}
\newcommand{\lfmbias}{\textsc{LFM2b-DB}\xspace}
\newcommand{\lfmbiasshort}{\textsc{LFM2b-DB}}

\newcommand{\LsRec}{\Ls^{\text{rec}}}
\newcommand{\LsAdv}{\Ls^{\text{adv}}}
\newcommand{\LsMult}{\Ls^{\text{MULT}}}
\newcommand{\LsKL}{\Ls^{\text{KL}}}
\newcommand{\LsCE}{\Ls^{\text{CE}}}

\def\Tabref#1{Table~\ref{#1}}

\copyrightyear{2022}
\acmYear{2022}
\setcopyright{acmcopyright}
\acmConference[SIGIR '22]{Proceedings of the 45th International ACM SIGIR Conference on Research and Development in Information Retrieval}{July 11--15, 2022}{Madrid, Spain.}
\acmBooktitle{Proceedings of the 45th International ACM SIGIR Conference on Research and Development in Information Retrieval (SIGIR '22), July 11--15, 2022, Madrid, Spain}
\acmPrice{15.00}
\acmISBN{978-1-4503-8732-3/22/07}
\acmDOI{10.1145/3477495.3531820}

\begin{document}
\fancyhead{}
\title{Unlearning Protected User Attributes in Recommendations with Adversarial Training}

\author{Christian Ganh{\"{o}}r}
\email{christian.ganhoer@jku.at}
\orcid{0000-0003-1850-2626}
\affiliation{%
  \institution{Johannes Kepler University Linz}
  \country{Austria}
}

\author{David Penz}
\email{david.penz@jku.at}
\orcid{0000-0002-7168-8098}
\affiliation{%
  \institution{Johannes Kepler University Linz and TU Wien}
  \country{Austria}
}

\author{Navid Rekabsaz}
\email{navid.rekabsaz@jku.at}
\orcid{0000-0001-5764-8738}
\affiliation{%
  \institution{Johannes Kepler University Linz and Linz Institute of Technology}
  \country{Austria}
}

\author{Oleg Lesota}
\email{oleg.lesota@jku.at}
\orcid{0000-0002-8321-6565}
\affiliation{%
  \institution{Johannes Kepler University Linz and Linz Institute of Technology}
  \country{Austria}
}

\author{Markus Schedl}
\email{markus.schedl@jku.at}
\orcid{0000-0003-1706-3406}
\affiliation{%
  \institution{Johannes Kepler University Linz and Linz Institute of Technology}
  \country{Austria}
}

%%
%% By default, the full list of authors will be used in the page
%% headers. Often, this list is too long, and will overlap
%% other information printed in the page headers. This command allows
%% the author to define a more concise list
%% of authors' names for this purpose.
\renewcommand{\shortauthors}{Ganh{\"{o}}r et al.}

%%
%% The abstract is a short summary of the work to be presented in the
%% article.
\begin{abstract}
Collaborative filtering algorithms capture underlying consumption patterns, including the ones specific to particular demographics or protected information of users, \eg gender, race, and location. 
%Recent studies have shown that recommendation systems (RSs), when learning from interaction data, also encode the societal biases regarding users' protected attributes (\eg gender and race) in their models. 
These encoded biases can influence the decision of a recommendation system (RS) towards further separation of the contents provided to various demographic subgroups, and raise privacy concerns regarding the disclosure of users' protected attributes.
%Such encoded biases can be harmful if they result in delivering inferior recommendations to certain user groups. 
In this work, we investigate the possibility and challenges of removing specific protected information of users from the learned interaction representations of a RS algorithm, while maintaining its effectiveness.
%In this work, we take a step towards mitigating such biases in recommendation models, by proposing an adversarial approach that learns invariant interaction representations with respect to protected attributes. 
Specifically, we incorporate adversarial training into the state-of-the-art \multvae architecture, resulting in a novel model, \emph{Adversarial Variational Auto-Encoder with Multinomial Likelihood} (\model), which aims at removing the implicit information of protected attributes while preserving recommendation performance. We conduct experiments on the MovieLens-1M and LFM-2b-DemoBias datasets, and evaluate the effectiveness of the bias mitigation method based on the inability of external attackers in revealing the users' gender information from the model. Comparing with baseline \multvae, the results show that \model, with marginal deterioration in performance (\wrt \ndcg and \recall), largely mitigates inherent biases in the model on both datasets.

% models in terms of \ndcg and \recall metrics for performance, and Balanced Accuracy for bias mitigation. 
% TODO: future: what we show and what directions are opened that way
% TODO: Reevaluate for recall
%\vspace{-2mm}
\end{abstract}

%%
%% The code below is generated by the tool at http://dl.acm.org/ccs.cfm.
%% Please copy and paste the code instead of the example below.
%%


%%
%% Keywords. The author(s) should pick words that accurately describe
%% the work being presented. Separate the keywords with commas.

%% A "teaser" image appears between the author and affiliation
%% information and the body of the document, and typically spans the
%% page.
%\begin{teaserfigure}
%\end{teaserfigure}

%%
%% This command processes the author and affiliation and title
%% information and builds the first part of the formatted document.

\maketitle

\begin{figure}[h]
\centering

\subfloat[\multvae]{
\includegraphics[width=0.49\linewidth]{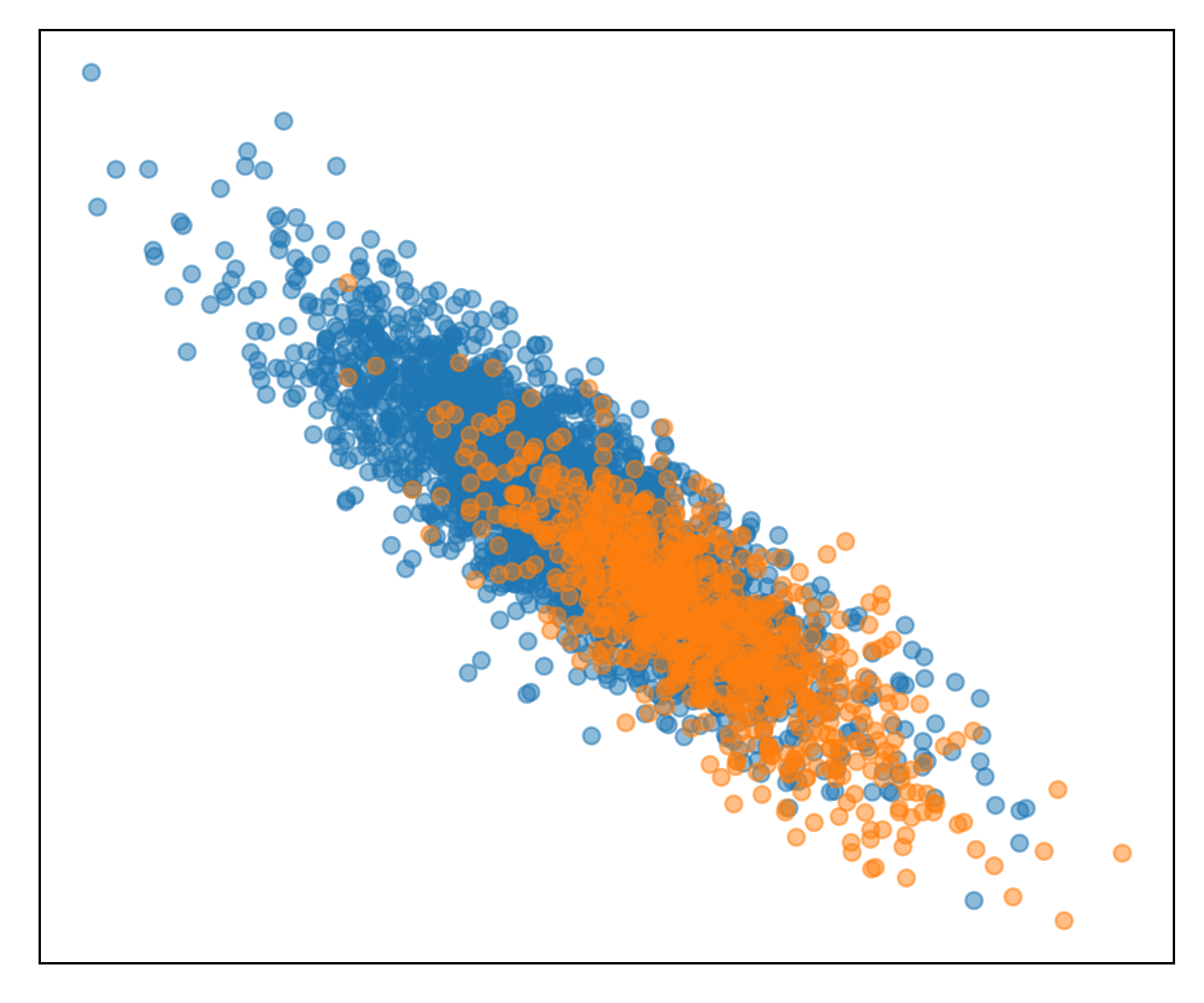}\label{fig:projection:multvae}}
%\hfill
\subfloat[\model]{
\includegraphics[width=0.49\linewidth]{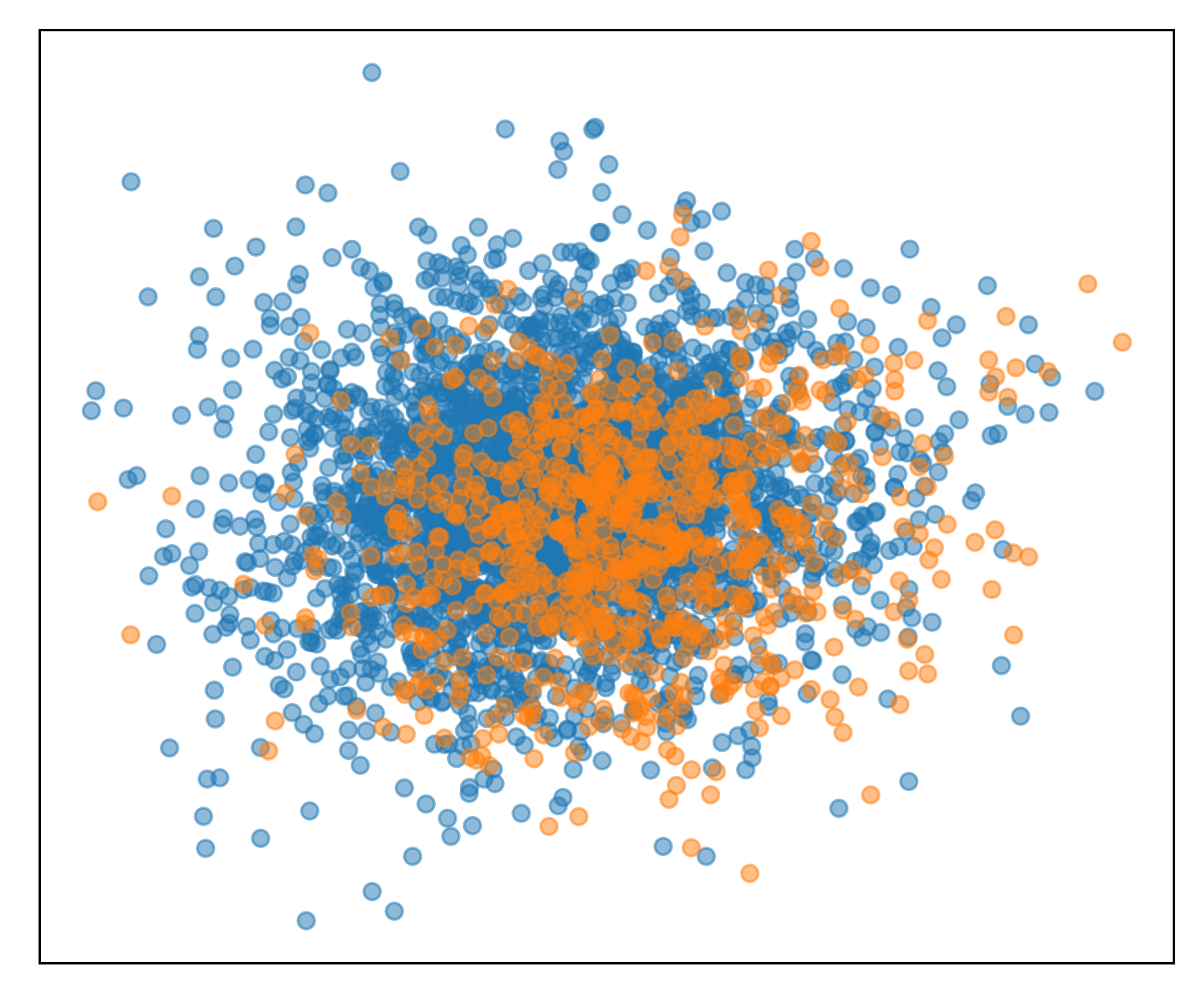}\label{fig:projection:model}
}
\label{fig:projection}
%\vspace{-2mm}
\caption{Output of an attacker network aiming to infer users' genders from the latent embeddings of the \multvae and \model models trained on LFM-2b-DemoBias~\cite{melchiorre_gender_fairness} dataset. The blue and orange markers correspond to male and female users, respectively.}
%\vspace{-4mm}
\end{figure}

\vspace{-2mm}
\section{Introduction}

% abstract + intro 1 page
In recommender systems (RSs), collaborative filtering algorithms provide recommendations for users (consumers), primarily based on the collected user-item interactions, \eg through listening to music tracks or watching movies. Among these algorithms, \multvae~\cite{liang_mult_vae} learns to recommend items through decoding the variational encoding of user interaction vectors and has shown top results among a variety of deep neural network approaches~\cite{DBLP:conf/recsys/DacremaCJ19}. While the interaction data does not explicitly contain information about protected user attributes such as gender, race, or age, a model may still encode sensitive information in its latent embeddings. This is depicted in Figure~\ref{fig:projection:multvae}, as the points regarding male and female users in a trained \multvae model form fairly separated clusters of users according to their genders. %As discussed in previous work, 
These encoded biases in models can lead to strengthening ``filter bubbles'' based on the demographics of users~\cite{10.1145/3460231.3474244,ekstrand2018exploring,elahi2021towards,bauer2019allowing}, and to intensifying the existing societal biases in data, thereby increasing unfairness of the RS~\cite{espin2022inequality,melchiorre_gender_fairness,rekabsaz_societal_biases_adv_bert,rekabsaz2020do}. They can also raise privacy concerns regarding the disclosure of sensitive information from the recommendations or model parameters~\cite{beigi_privacy_aware_recommendations,beigi_privacy_conserving_text,weinsberg_blurme_sensitive_information_in_history}.

%Such unwanted encoding of protected attributes can lead to privacy concerns~\cite{beigi_privacy_aware_recommendations,beigi_privacy_conserving_text} and \emph{un}fairness in recommendation~\cite{melchiorre_gender_fairness, lesota_popularity_bias}, particularly impacting the minority user groups, either through being overly underrepresented (representational harm) or being provided with significantly lower quality recommendations (allocational harm).

% which can then be used to generate recommendations with better levels of personalization~\cite{weinsberg_blurme_sensitive_information_in_history}

We approach this issue by proposing \model, a novel bias-aware recommendation model which enhances \multvae with adversarial training to reduce encoded biases. The \model model, while learning to provide effective recommendations, simultaneously forces its latent embeddings to become invariant with respect to a given protected attribute of the consumers. This results in reducing the distinguishability of the sub-populations in the model (as shown in Figure~\ref{fig:projection:model}), hence making the recommendation ``blind'' to the protected attribute while maintaining the model's recommendation performance. We particularly adopt \multvae, as it achieved top results among a variety of different deep neural network based approaches~\cite{DBLP:conf/recsys/DacremaCJ19}.

To assess the merits of our approach regarding both bias mitigation and recommendation performance, we conduct a set of experiments on the MovieLens-1m~\cite{harper_movielens_2015} and LFM-2b-DemoBias~\cite{melchiorre_gender_fairness,schedl2022lfm} datasets covering the domains of movies and music, respectively. We focus on gender as the protected attribute and evaluate the accuracy and balanced accuracy of an attacker network to quantify the effect of bias mitigation. Moreover, we assess the models’ recommendation performance via \ndcg and \recall. %We observe that 
\model successfully reduces inherent gender bias, whilst marginally decreasing performance mainly caused by the challenges imposed during model selection.

\paragraph{Brief Review of Related Work}
As surveyed by~\citet{DBLP:journals/csur/DeldjooNM21}, adversarial training in combination with latent factor recommendation algorithms is investigated for various purposes by a few recent studies. In particular, \citet{beigi_privacy_aware_recommendations} propose a novel model based on Bayesian Personalized Ranking (BPR), which uses attacker networks to increase the model's privacy. In this model, the attacker networks aim to infer sensitive user information by looking at the output recommendations of the network, and the whole model is optimized such that no sensitive information can be inferred from the recommendations. Similarly, \citet{zhang_MitigatingUnwantedBiases_2018} intend to mitigate biases of classifiers by utilizing adversarial networks, resulting in reducing the leakage of sensitive user attributes into the model predictions. In contrast to these studies, our proposed model aims to remove implicitly encoded sensitive information from its latent space rather than the output space. Moreover, unlike some approaches \cite{wu_LearningFairRepresentations_2021, li_PersonalizedFairnessBased_2021} that apply filtering layers on top of their user embeddings to drop unwanted information, \model is trained with the objective that the information of the protected attributes is removed from the model in the first place.
Concerning bias mitigation in RSs, \citet{DBLP:conf/sigir/ZhuWC20} introduce the debiased personalized ranking model, in which the adversarial training aims to identify which item group, such as movie genre in the movie domain, the recommendation belongs to. This information is subsequently removed to mitigate \textit{item popularity bias}. In contrast to this work, we study bias mitigation from the consumer side. 
More recently, based on adversarial training, \citet{DBLP:conf/aaai/WuWWH021} explore the mitigation of consumer bias in news recommendation, and several recent studies\cite{rekabsaz_societal_biases_adv_bert,zerveas2022mitigating,krieg2022do,krieg2022grep} approach fairness in the representation of gender-related documents in information retrieval. Our work extends these studies by introducing a novel bias-aware recommendation model based on variational autoencoders.

The paper is structured as follows: We introduce \model in \Secref{section:adv-mult-vae}. In \Secref{section:experiment-setup}, we present our experimental setup and the datasets and metrics we use to evaluate our approach. \Secref{section:results-discussion} provides an analysis of our results, which we extend by open challenges and limitations in  \Secref{section:challenges-limitations}. Finally, we conclude this work in  \Secref{section:conclusions}. Our code together with all resources is available at \textbf{\url{https://github.com/CPJKU/adv-multvae}}\label{section:introduction}

\section{Adversarial \multvae}\label{section:adv-mult-vae}
\begin{figure}
    \begin{center}
        \includegraphics[width=1\linewidth]{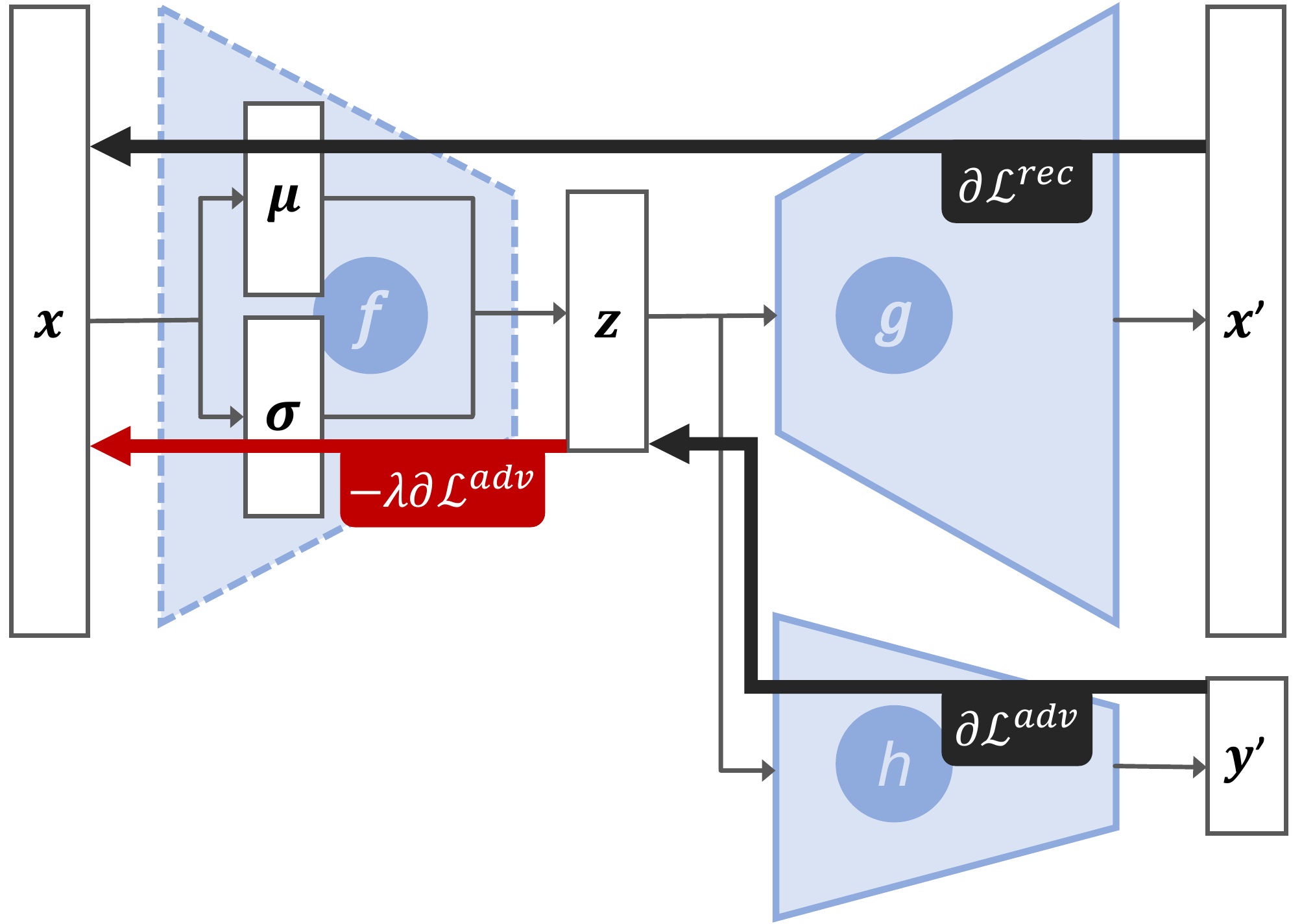}
        \caption{Outline of \model. The bold arrows show the flow of gradients during backward pass, where the red color indicates the reversed gradient for learning latent embeddings ($\vz$) invariant to the protected attribute ($\vy$).}
        \label{fig:model-outline}
    \end{center}
    %\vspace{-6mm}
\end{figure}

% half page
In this section, we describe the architecture of our \emph{Adversarial Variational Auto-Encoder with Multinomial Likelihood} (\model) model. %, an extension of the \multvae model~\cite{liang_mult_vae} with an adversarial component, which aims to reduce the encoded information of users' sensitive attributes. %In the following, 
We first provide an overview of the baseline \multvae, followed by explaining our adversarial extension. We finally describe the procedure of adversarial attacking used to assess the effectiveness of bias mitigation. \Figref{fig:model-outline} depicts the outline of the proposed \model model.

%This is important, as otherwise, recommendations may be partially based on \eg gender information~\cite{weinsberg_blurme_sensitive_information_in_history}. 

\paragraph{\multvae}
The \multvae model consists of two parts: First, the encoder network $f(\cdot)$ receives the input vector $\vx$ containing the interaction data of a user and infers a low-dimensional latent distribution. Considering a standard Gaussian distribution as the prior ($\mathcal{N}(0,I)$) and using the \textit{reparameterization trick}~\cite{journals/corr/KingmaW13}, this distribution is characterized by $\vmu$ and $\vsigma$ learnable vectors, from which the latent vector $\vz$ is sampled.~\footnote{In short, reparametrization trick allows sampling the random variable $\vz$ by reparameterizing the sampling process with an auxiliary stochastic variable, thereby maintaining the ability to perform back-propagation on $\vmu$ and $\vsigma$.} The second part is the decoder network $g(\cdot)$, which aims to reconstruct the original input $\vx$ from the latent vector $\vz$ by predicting $\vx'$. We refer to the loss function of \multvae as $\LsRec(\vx)$, defined below:
\begin{align}
    \LsRec(\vx) = \LsMult(g(\vz), \vx) - \beta \LsKL(\mathcal{N}(\vmu,\vsigma), \mathcal{N}(0,I)) 
\end{align}
where $\LsMult$ is the input reconstruction loss, and $\LsKL$ is the regularization loss aiming to keep the latent distribution of the encoder close to the prior, whose influence is adjusted by the hyperparameter $\beta$. We refer to~\citet{liang_mult_vae} for more details. 

%which in principle consists of a reconstruction loss and a term to impose the prior distribution (for more details see \citet{liang_mult_vae}).

%This reduction of dimensions leads to a loss of information of low importance (such as noise).

%
%Following Inspired by previous work on adversarial training~\cite{rekabsaz_societal_biases_adv_bert, ganing_gradient_reversal,elazar-goldberg-2018-adversarial,xie2018controllable},

%\paragraph{Adversarial \multvae} Our proposed
\paragraph{\model} 
Our proposed model extends \multvae with an adversarial network, referred to as $h(\cdot)$. The adversarial network is added as an extra head over the latent vector and aims to predict from latent vector $\vz$ a specific protected attribute of the user. %a specific protected attribute of the user from $\vz$ as the given input. 
$h(\cdot)$ is typically a feedforward network, which is optimized with respect to the $\vy$ vector containing the user's protected attribute as classification labels. The training process of \model aims to simultaneously remove the information of the protected attribute from $\vz$, and maintain recommendation performance. To this end, the loss of the model is defined as the following min-max problem:
\begin{equation}
\begin{aligned}
    \argmin_{f, g} \argmax_{h} \LsRec(\vx) - \LsAdv(\vx, \vy) %\\
    %\text{i.e.,} \quad \LsAdv(\vx, \vy) = \LsCE(h(\vz), \vy)
\end{aligned}
\label{eq:advloss1}
\end{equation}
where the loss of the adversarial network $\LsAdv$ is defined as the cross-entropy loss ($\LsCE$) between the predicted and actual value of the protected attribute: $\LsAdv(\vx, \vy) = \LsCE(h(\vz), \vy)$. 
%In fact, the loss defined in Eq.~\ref{eq:advloss1} aims to maximize the prediction ability of $h(\cdot)$ by decreasing $\LsAdv$ via $\argmax_{h}$ when $\vz$ is given, while it minimizes the encoded information in $\vz$ concerning the protected attribute by increasing $\LsAdv$ via $\argmin_{f}$. % and provides effective recommendations. 
In fact, the loss defined in Eq.~\ref{eq:advloss1} aims to maximize the prediction ability of $h(\cdot)$ to discover all sensitive information when $\vz$ is given, while it minimizes the encoded information in $\vz$ concerning the protected attribute.

%, namely improving the recommendation quality, and lowering the adversarial performance

Considering the well-known complexities of optimizing min-max loss function~\cite{NIPS2014_5ca3e9b1}, following previous work~\cite{rekabsaz_societal_biases_adv_bert,elazar-goldberg-2018-adversarial,xie2018controllable}, we add a gradient reversal layer $grl(\cdot)$~\cite{ganing_gradient_reversal} between $\vz$ and the adversarial network $h(\cdot)$. During training, $grl(\cdot)$ acts as the identity function in the forward pass, while it scales the calculated gradient by $-\lambda$ in the backward pass. The $grl(\cdot)$ network does not have any effect on the model at inference time. We refer to the parameter $\lambda$ as \textit{gradient reversal scaling}. By employing $grl(\cdot)$ in the model, the overall loss in Eq.~\ref{eq:advloss1} can now be reformulated to a standard risk minimization setting: %, as shown below:
\begin{equation}
\begin{aligned}
    \argmin_{f, g, h} \Ls = \LsRec(\vx) + \LsAdv(\vx, \vy), \\
    %\LsAdv(\vx, \vy) = \LsCE(h(grl(\vz), \vy)
    \LsAdv(\vx, \vy) = \LsCE(h(grl(\vz), \vy)
\end{aligned}
\end{equation}

This formulation enables optimizing the model through standard gradient-based loss minimization. %Overall, 
%The training of \model is expected to make the learned latent embeddings invariant (or less informative) with respect to the users' protected attribute.

%Let us note that although the adversarial network was defined only for a single protected attribute, this definition is extensible to multiple protected attributes by adding corresponding $\LsAdv$ terms to the overall loss $\Ls$. 

\paragraph{Adversarial Attacks.}
After training the model (whether \multvae or \model), we examine to which extent the information of the protected attribute remains in the model, \ie to which degree this information can still be recovered. To this end, once the training is complete, an attacker network~$h^{\text{atk}}(\cdot)$ is introduced to the model, which aims to predict the protected attribute~$\vy$ from the latent vector~$\vz$. Similar to $h(\cdot)$, the attacker $h^{\text{atk}}(\cdot)$ is defined as a feedforward network. During training the attacker, all model parameters remain unchanged (are frozen) and only the attacker parameters are updated. The prediction performance of the attacker -- %and in fact how much this performance is better than random predictions 
relative to a random predictor -- is used as a metric to quantify the degree %of existence 
of bias in the underlying model. 

% cite attacker paper, not sure what its name is... 
%it may not be able to delete all information to begin with (without strong increase of $\LsRec$). Moreover, the encoder may learn to actively the information from the adversary to reduce its loss. 

%\section{Recommendation Fairness Metric}\label{section:fairness-metric}
%\input{3-metric}

\section{Experiment Setup}\label{section:experiment-setup}
% about 1/2 page
%\dap{should we start with 1-2 sentences before jumping to Datasets? E.g. for the experiment setup we want to provide details on datasets used, evaluation metrics, and training procedure or something like that - for me it just looks a bit ugly to directly start with a subsection}

%, namely the datasets, utility and bias metrics, and finally the training procedure

In this section, we describe the setup of our experiments. To ensure reproducibility, %that the experiments are fully reproducible, 
our dataset splits, code and hyperparameters are available at \textbf{\url{https://github.com/CPJKU/adv-multvae}}.

%\vspace{-1mm}
\paragraph{Datasets.}
We evaluate our approach on two standardized datasets containing user-item interactions as well as partial demographic information of their users: \textbf{(1) \movielens~\cite{harper_movielens_2015}}\footnote{\url{https://grouplens.org/datasets/movielens/1m}} contains ratings of users on movies as well as the users' gender, age, and occupation information. We binarize the interactions by setting the values of the rated items to one, and the rest to zero. %, to generate implicit feedback data.
%We binarize the ratings to use them as implicit feedback data.
Finally, we only keep the users that rated at least 5 movies, and the movies with at least 5 user interactions; \textbf{(2) \lfmbias~\cite{melchiorre_gender_fairness}\footnote{\url{http://www.cp.jku.at/datasets/LFM-2b}}} %the \lfm Demographic Bias dataset 
is a subset of the \lfm dataset, which provides a collection of music listening records of users, for whom partial demographic information (gender, age, country) is available. We follow the same experiment setting as in \citet{melchiorre_gender_fairness}. In particular, we only keep the user-item interactions with a play count of at least 2, and binarize the interactions. %, filtering possible unintended interactions (misclicks). We binarize the interactions. 
Moreover, for computational reasons, 100,000 tracks are randomly sampled from the data. Finally, we keep only the users with at least 5 track interactions, and the tracks that are listened to at least 5 times.
%Moreover, we keep only the users with at least 5 track interactions, and the tracks that are listened to at least 5 times. Finally, for computational reasons, 100,000 tracks are randomly sampled from the remaining data. %\footnote{\Tabref{tab:dataset} reports slightly fewer items as we ensure that the minimum number of interactions for users and items is preserved.} 
The statistics of the datasets are reported in \Tabref{tab:dataset}. With both datasets, we focus on the users' gender as the protected attribute for our experiments.\footnote{The provided gender information of the users in the datasets are limited to female and male, neglecting the more nuanced definition of genders. Despite this limitation, the introduced model is generic and can be applied to non-binary settings too.} 

\begin{table}[t]
    \begin{center}

\begin{tabular}{llrrr}
\toprule
Dataset  &  \multicolumn{2}{c}{Users} & Items & Interactions \\
\midrule
\multirow{3}{*}{\movielens} & All & 6,040     & \multirow{3}{*}{3,416}  & 999,611      \\
& Male   & 4,331 & & 753,313      \\
& Female   & 1,709     & & 246,298      \\
\midrule
\multirow{3}{*}{\lfmbias} & All & 19,972    & \multirow{3}{*}{99,639} & 2,829,503    \\
 & Male & 15,557 &  & 2,385,427    \\
 & Female & 4,415 & & 444,076   \\ \bottomrule
\end{tabular}
        \caption{Statistics of the datasets used in our experiments.}
        \label{tab:dataset}
    \end{center}
    %\vspace{-8mm}
\end{table}
 
%\vspace{-1mm}
\paragraph{Data Splits.} Following the exact setting of \citet{melchiorre_gender_fairness}, we apply a user split strategy~\cite{meng_data_splitting_strategies}. In this setting, the users (and their corresponding interactions) are split into 5 folds for cross-validation, where 3 folds make up the training set, and 1 fold each makes up the validation and test set. For the training set, we further perform random upsampling of female users (as the minority group) to achieve an balanced dataset, which supports bias mitigation in models~\cite{melchiorre_gender_fairness}. 
For validation and testing, the interactions in each set are further split: 80\% are used as model input, the remaining 20\% for calculating the evaluation metrics.
%For validation and testing, the interactions in each set are further split according to an 80-20 proportion: the first part is used as the model input, and the second part is used for calculating the evaluation metrics.
 
%\vspace{-1mm}
\paragraph{Evaluation.} We use two popular recommendation performance metrics: \recallk, namely the fraction of relevant items in the top $k$ recommended items, and \ndcgk, which weights the relevance of the top $k$ recommended items based on their ranking positions. As common, we set the cut-off threshold $k$ to 10. %in our experiment. 
Additionally, we measure the effectiveness of the models in terms of bias mitigation using the accuracy (\acc) and balanced accuracy (\bacc) of the attacker when predicting the users' gender. We use \bacc as a proper metric in imbalanced classification settings~\cite{brodersen_balanced_accuracy}. It reports the average recall per class (female/male) where a value of $0.50$ indicates a fully debiased network. We report the performance and the bias mitigation results as the average over all test sets' results across cross-validation folds. We test the statistical significance of the differences of the performance metrics using the Wilcoxon signed-rank test~\cite{DBLP:reference/stat/ReyN11} with a confidence level of 95\%.% While we cannot determine whether the decrease in \bacc is significant\footnote{In theory it is possible, however, as \bacc is computed over the whole test set, i.e., is only a single value per set, we would need to repeat the experiments many times to gather sufficient data.}, we can however test if the performance of the attackers on the model with and without adversarial training is significantly different via a McNemar's test~\cite{McNemar1947}. We use a confidence level of 95\% for both tests to determine significance.

%One should note that although \bacc values close to 1 are usually regarded as good, in our bias use-case, lower values are preferred as they imply a low amount of sensitive information that is still available in the network.

%since we want to remove sensitive information from the network, low values are preferred, \eg for binary classification, a value close to 0.5.
%\cg{State \rrb as a one of the bias metrics we consider for evaluation}

%\vspace{-1mm}
\paragraph{Models and Training.} 
We train the \multvae and \model models for 100 and 200 epochs on \lfmbias and \movielens, respectively. For \multvae, we select the best performing model across the training epochs based on the validation \ndcg results. We refer to this model as \textbf{\multvaebest}. For \textbf{\model}, we conduct model selection based on the \bacc results of the adversarial network, mainly resulting in the selection of the model at the very last steps of training. To have a comparable setting between \multvae and \model, we also report the results of \multvae when the selected model is at the last epoch, denoted as \textbf{\multvaelast}. Additionally, we perform a hyperparameter search over embedding size, parameter $\beta$, various dropouts, learning rate, weight decay, and gradient reversal scaling $\lambda$. In each setting, the best performing model on \ndcg on the validation set is chosen and evaluated on the test set. The range of hyperparameters and the ones used for the final models are available together with our published code. 

%to obtain a model with stable training and which does not overfit on the training set. (2) For the found configuration, we then tune a new \model model with gradient scaling $\lambda > 0$, again ensuring proper training. Finally, we perform a search over a wide range of gradient scalings to determine the parameters influence on model utility and bias. 
%Furthermore, we attack each trained model as described in the model outline in \Secref{section:adv-mult-vae} by training an attacker network on the latent space encoded trainset and validate and evaluate analogously to the adversarial training (\bacc is used to report the attackers performance).

%\cg{need statement about non-binary gender somewhere}

%\vspace{-1mm}
\section{Results and Analyses}\label{section:results-discussion}
% \cg{as the ranking bias is really important for us, we validate @10} -- moved to Evaluation Metrics?

\begin{table}[t]
\begin{center}
    \begin{tabular}{llll|ll}
\toprule
\multirow{2}{*}{Dataset} & \multirow{2}{*}{Model} & \multicolumn{2}{c|}{Bias$\downarrow$}  & \multicolumn{2}{c}{Performance$\uparrow$} \\
& & \multicolumn{1}{c}{\acc} & \multicolumn{1}{c|}{\bacc}  & \multicolumn{1}{c}{\ndcg} & \multicolumn{1}{c}{\Recall}\\
\midrule
\multirow{3}{*}{\movielens} & \multvaebest & 0.692  & 0.707  & \textbf{0.621} & \textbf{0.596} \\
& \multvaelast & 0.699 & 0.693  & 0.591$\dagger$ & 0.566$\dagger$ \\
& \model & \textbf{0.565} & \textbf{0.572} & 0.593$\dagger$ & 0.569$\dagger$  \\
\midrule
\multirow{3}{*}{\lfmbiasshort} & \multvaebest & 0.703 & 0.717 & \textbf{0.211} & \textbf{0.192} \\
& \multvaelast & 0.709 & 0.717 & 0.206$\dagger$ & 0.189$\dagger$\\
& \model & \textbf{0.631} & \textbf{0.609} & 0.206$\dagger$ & 0.189$\dagger$ \\
\bottomrule
\end{tabular}
    \caption{The results of bias mitigation in terms of accuracy and balanced accuracy of adversarial attackers (lower values indicate less bias), as well as recommendation performance (\ndcg and \recall). The best results are shown in bold. The sign $\dagger$ indicates a significant decrease in performance metrics in comparison to \multvaebest.\protect\footnotemark
    }
    %\protect\footnotemark %On the other hand, the sign $\ddagger$ indicates a significantly different performance (\acc and \bacc) of the attackers in comparison to \multvaebest.
    \label{tab:results}
    %\vspace{-8mm}
\end{center}
\end{table}

\footnotetext{We omit testing for significance on \acc and \bacc as they are both metrics calculated on the whole test set, which would require running the experiments many times to gather sufficient data. However, a McNemar's test~\cite{McNemar1947}, which we use to determine whether the attackers on the different models achieve different performances, signals a significant difference of the results ($p<0.05$).}

%\Tabref{tab:results} reports the results of the bias mitigation and utility metrics. We report the best performing results in bold, and indicate the significant decrease in the results of utility metrics in comparison with \multvaebest with the $\dagger$ sign. %In the rest of this section, we first detail the results of bias mitigation, then analyze the effect of adversarial training on utility metrics, and finally discuss the possible limitations of the study.

%\cg{Reviewers suggested to perform more in-depth analysis on which biases are actually removed. Average of metrics over all users may not give enough insight. We should also consider "diversity" and perhaps "popularity bias" to better explain the result of adversarial training.}
%\subsection{Bias Mitigation Results}
\Tabref{tab:results} reports the results of the bias mitigation and recommendation performance metrics. We report the best performing results in bold, and indicate the significant decrease in the results of performance metrics in comparison with \multvaebest with the $\dagger$ sign. 

Considering the balanced accuracy results of the standard \multvae model in \Tabref{tab:results}, we observe that the gender of the users can be identified from their consumption patterns with considerably high values, namely with $\sim\!0.71$ \bacc in \movielens and $\sim\!0.72$ in \lfmbias, in comparison to $0.5$ of a bias-free model. This confirms the existence of gender bias in \multvae and consequently in its provided recommendations, as also similarly reported in previous studies~\cite{melchiorre_gender_fairness,ekstrand2018all,schedl2015influence}. Looking at the results of \model, we observe a significant decrease in the balanced accuracy of attackers, indicating the effectiveness of \model for mitigating encoded societal biases. Despite the large decreases, we should note that the predictions are not yet fully random (\bacc $>\!0.5$), indicating that the model still contains considerable biases.

\begin{figure}
    \begin{center}
        \includegraphics[width=0.7\linewidth]{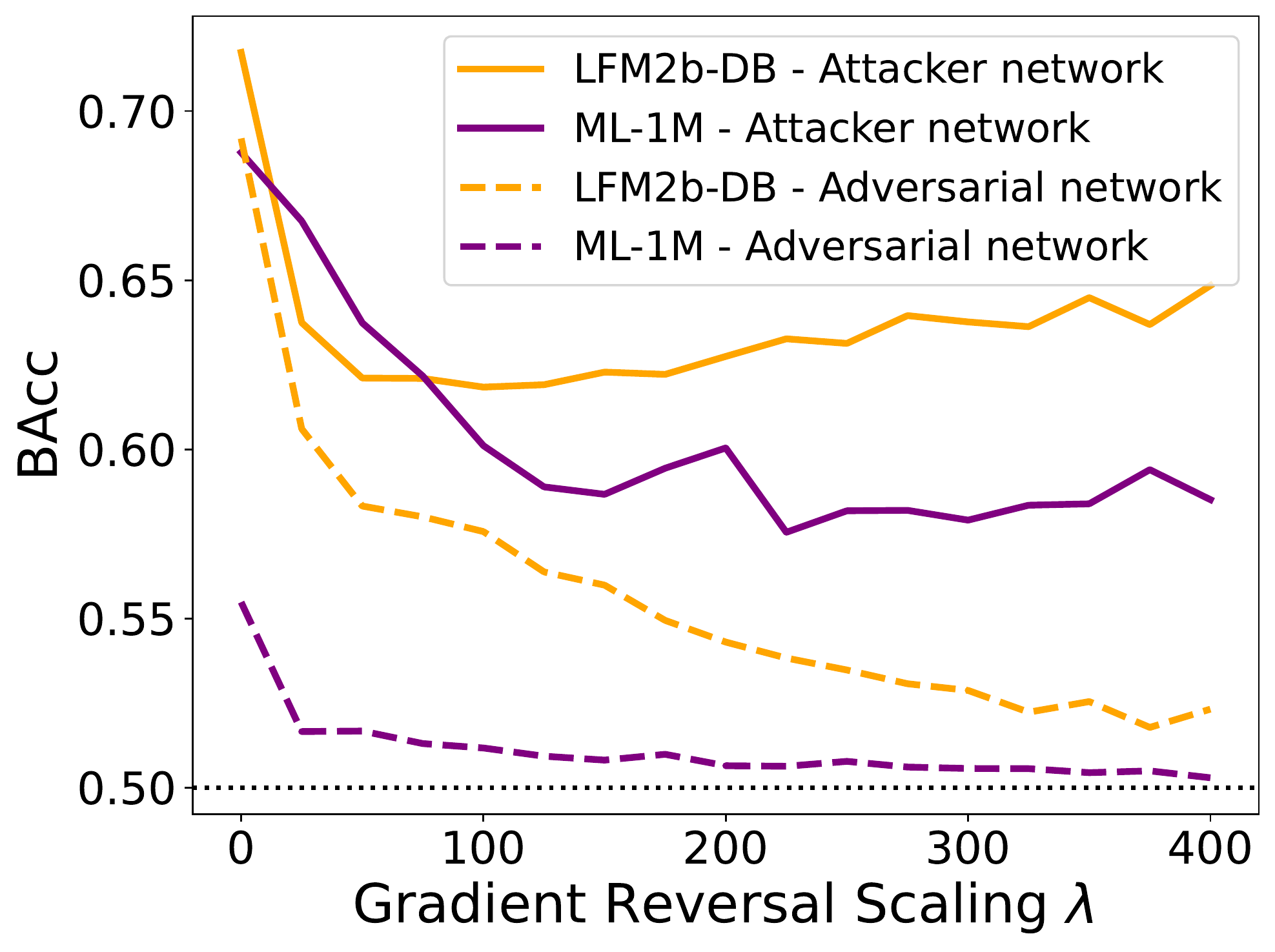}
        \caption{Balanced accuracy of adversarial and attacker networks over a range of gradient reversal scalings $\lambda$.}
        \label{fig:grad-scaling-influence}
    \end{center}
    %\vspace{-5mm}
\end{figure}

\begin{figure}[tb]
\centering
\subfloat[\movielens]{\includegraphics[width=0.47\linewidth]{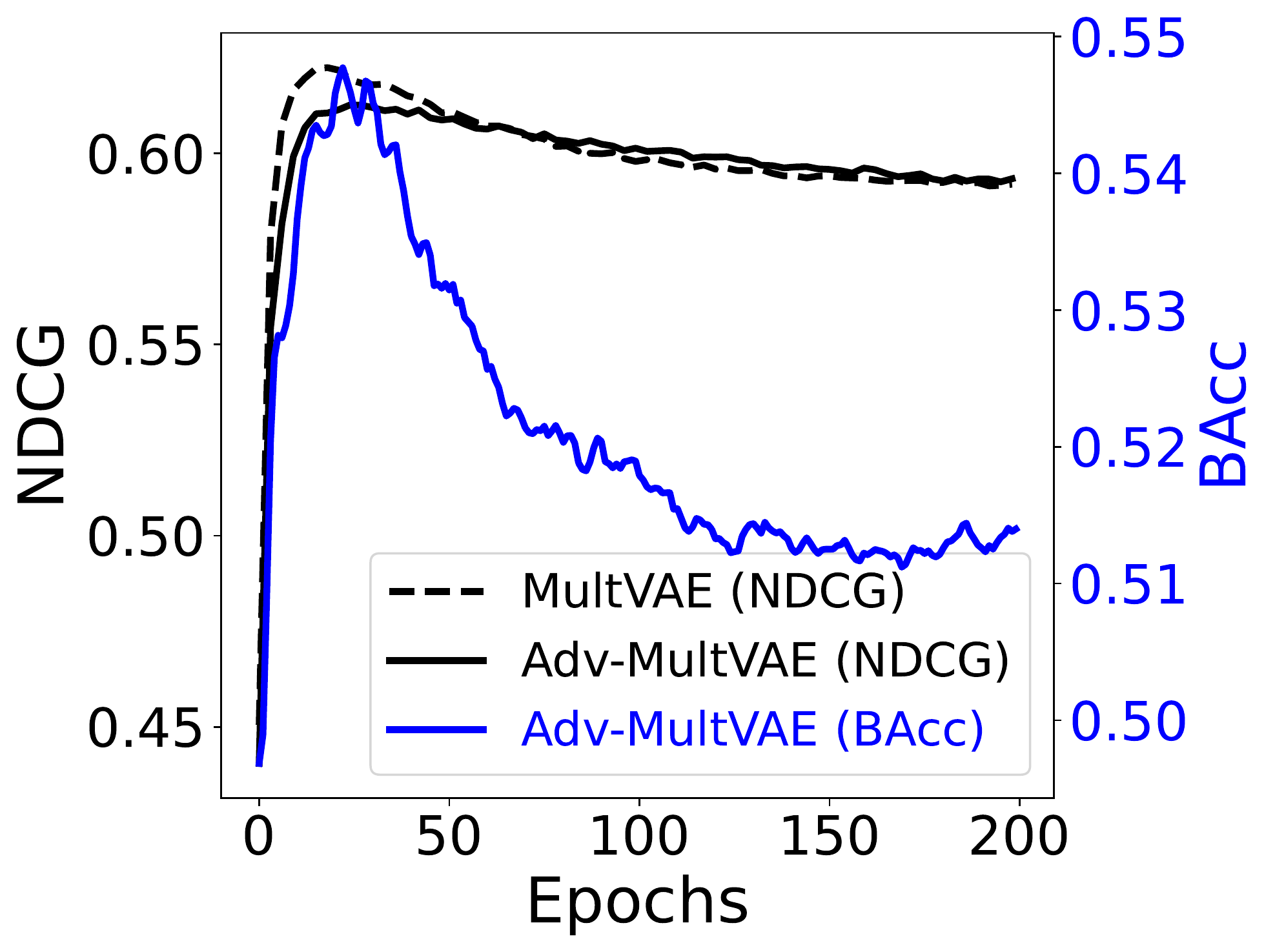}\label{fig:training:ml}}
\hspace{3mm}
\subfloat[\lfmbias]{\includegraphics[width=0.47\linewidth]{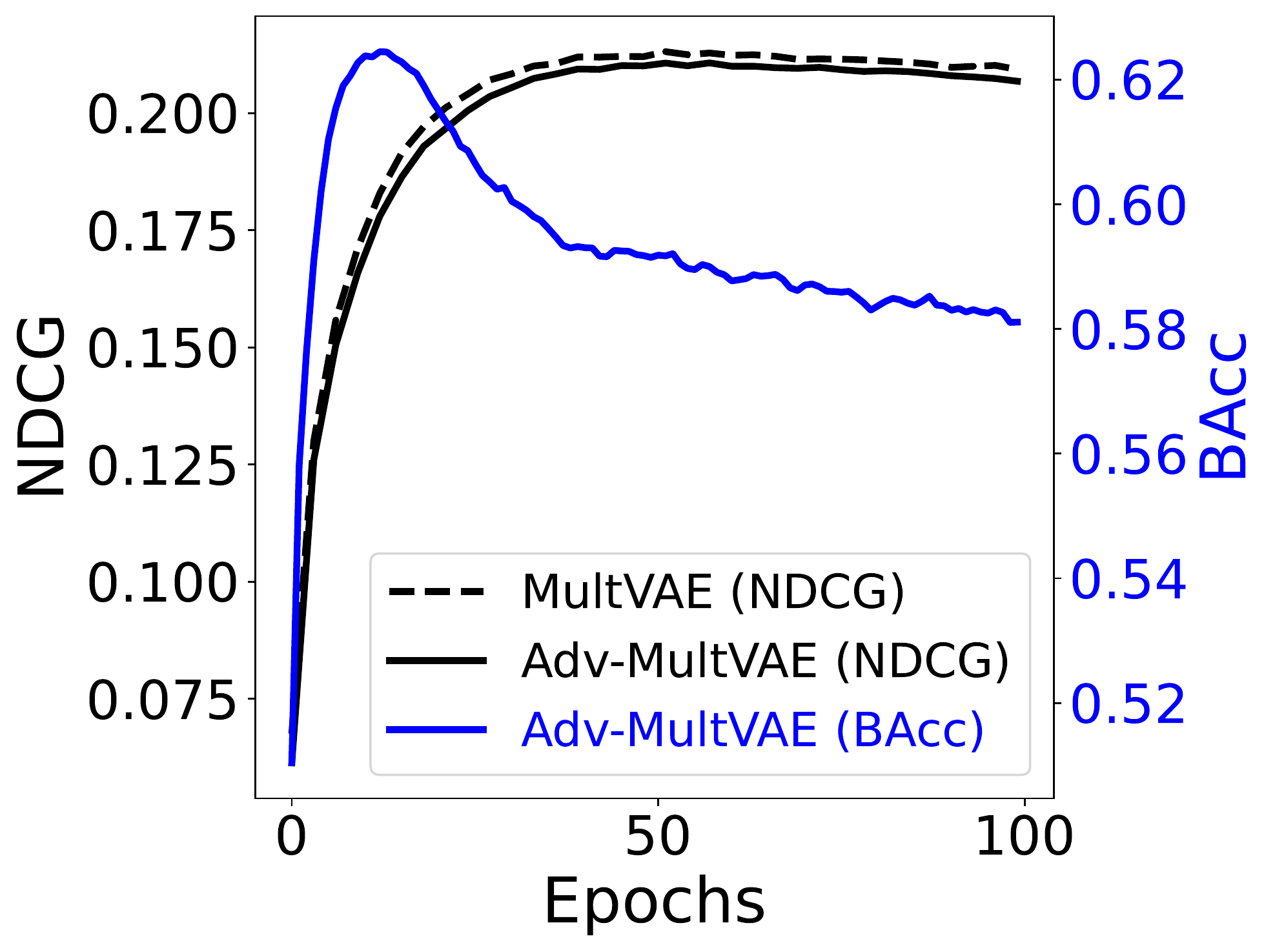}\label{fig:training:lfm}}
\caption{Performance evaluation results on the validation sets during the training of models for both \model and \multvae (NDCG on the left y-axes). The \bacc of the adversarial network in \model is shown based on the scale of the right y-axes.}
\label{fig:training}
%\vspace{-6mm}
\end{figure}

%In the plot, the blue lines correspond to the models evaluated on the \movielens dataset, while the green lines correspond to the \lfmbias dataset. Furthermore, the dashed lines report the adversaries \bacc, whereas the solid lines report the attackers \bacc. As already previously described in \Secref{section:adv-mult-vae}, we consider the former less important.

Let us have a closer look at the impact of gradient reversal scaling $\lambda$ on the bias mitigation method. \Figref{fig:grad-scaling-influence} shows the balanced accuracy of the attackers as well as adversarial networks for both datasets over a range of $\lambda$ values. As shown, by increasing $\lambda$, the \bacc of the adversarial networks generally decrease and saturate at some point. In particular, the adversarial network of the model trained on \movielens much earlier reaches the \bacc of close to 0.5. However, the attackers' \bacc are consistently higher than the ones of adversaries, indicating that the adversarial network could not discover and remove all sensitive information.
\vspace{-1mm}
\paragraph{Impact of adversarial training on performance}
Comparing the performance evaluation results of \model to that of \multvaebest in \Tabref{tab:results}, we observe a significant drop in \ndcg and \recall on both datasets. At first, this drop might be seen as the \emph{cost} of applying adversarial bias mitigation. However, we should consider that the model selection of \model is done based on the lowest value of \bacc rather than the highest \ndcg (as in \multvae). %, which might not be an ideal choice for utility metrics.
%Therefore, 
This might not be an ideal model selection criterion as it ignores a potential trade-off between \bacc and \ndcg, and under-emphasizes recommendation performance of \model.

To better understand this, \Figref{fig:training} shows the \ndcg results on the validation sets of the two datasets for both, \model and \multvae models, over the training epochs. The figure also depicts \bacc of the adversarial network of \model. As shown, both models achieve higher values of \ndcg in early epochs, and by continuing the training, the performances slightly decrease (due to overfitting). However, \bacc in \model considerably decreases in later epochs, where \ndcg has already decreased due to the slight overfitting.

In order to examine the performances while factoring out the effect of model selection and thereby enabling a fairer comparison, we compare the recommendation results of \multvaelast with \model. As reported in \Tabref{tab:results}, we observe no significant differences in \ndcg and \recall between these two models, for both datasets.

%While we could only observe a significant drop in utility for one dataset as a trade-off for lower system bias, we expect drops similar to \movielens in other datasets. This requires recommender system providers to carefully weight the importance between utility and system bias for their user base.
%Moreover, we want to note that as reducing the system bias requires many more epochs of training than a good utility recommender, it is imperative that recommender part of the model does not overfit.

%\vspace{-1mm}
\section{Open Challenges and Limitations}\label{section:challenges-limitations}
The analysis in the previous section highlights the inherent challenge in training adversarial networks for bias mitigation. Since this method aims to simultaneously satisfy two objectives (increasing recommendation performance and decreasing bias), it is challenging to train and select the right model that optimizes both aspects.

%While \model indeed leads to a less biased system, it is also more challenging to train and evaluate. 

%they are also more challenging to train. Due to the competing nature of the two parts of the model, the recommender and the adversaries, where the latter is removing information the former would like to use, detecting problems in the training process is difficult. As an example, the \bacc reduction of the adversaries may either be caused by the getting rid of sensitive information, it may also be caused by overfitting on the train set.

We should also shed light on two limitations of our adversarial bias mitigation method for RSs. First, our adversarial approach aims to reduce the correlations in the model to the protected attribute based on the observed data. This approach, like other data-oriented bias mitigation methods, might lack effective generalization, particularly when the model is evaluated on other domains or out-of-distribution data. The second limitation of adversarial bias mitigation for RSs is that, while the method aims to make the recommendations agnostic to protected attributes, it does not directly account for the perception of users regarding the bias in the recommended results. 

%We envision addressing these limitation as future directions of this work, particularly by exploring the generalization aspects of the method, as well as the perception of end-users regarding the biases in recommendations.

%\vspace{-1mm}
\section{Conclusion and Future Directions}\label{section:conclusions}
This work addresses the challenge of mitigating societal biases in RSs from the user perspective. To this end, we extend the widely-used \multvae model with an adversarial component, and propose the novel \model architecture. %Our approach aims to decrease the encoded biases in the model and consequently also in the provided recommendations. 
Our approach aims to decrease the model bias in terms of latent information about the protected user attribute in the model and consequently also in the provided recommendations.
We conduct experiments on two datasets (\movielens and \lfmbias) and evaluate the amount of recoverable sensitive user information (gender in our experiments) from the models, as well as the models' recommendation accuracy. Our results show that the introduced \model model, despite yielding a marginal performance decrease which is mainly caused by the dilemma in model selection, provides a substantial reduction in the amount of encoded protected information, offering a bias- and privacy-aware alternative.

We envision addressing the limitation discussed in the previous section as future directions of this work, particularly by exploring the generalization aspects of the method, as well as the perception of end-users regarding the biases in recommendations. We would also like to gain further insights into which user groups specifically are affected by our approach.
Moreover, finding a balance between \bacc and \ndcg for optimization and model selection might be a possibility to mitigate the slight performance loss of \model.
%Moreover, we view the development of an interactive method to dynamically adjust the influence of user protected attributes as an interesting direction for further research.
%\nr{@Christian: expand it with other possible future directions to a full paragraph}

%\cg{We should perhaps also explain how our adversarial training approach works for non-binary attributes, as suggested by another reviewer. }

%For future work we consider extending \rrb to the non-binary context, allowing its use on a broader spectrum of sensitive attributes such as age and gender\footnote{without any simplifications}. Additionally, \rrb may also be used in comparing the fairness of different collaborative filtering approaches.

\begin{acks}
This work received financial support by the Austrian Science Fund (FWF): P33526 and DFH-23; and by the State of Upper Austria and the Federal Ministry of Education, Science, and Research, through grant LIT-2020-9-SEE-113 and LIT-2021-YOU-215.
\end{acks}

%% The next two lines define the bibliography style to be used, and
%% the bibliography file.
\bibliographystyle{ACM-Reference-Format}
\balance
\bibliography{references}

%%% -*-BibTeX-*-
%%% Do NOT edit. File created by BibTeX with style
%%% ACM-Reference-Format-Journals [18-Jan-2012].

\begin{thebibliography}{35}

%%% ====================================================================
%%% NOTE TO THE USER: you can override these defaults by providing
%%% customized versions of any of these macros before the \bibliography
%%% command.  Each of them MUST provide its own final punctuation,
%%% except for \shownote{}, \showDOI{}, and \showURL{}.  The latter two
%%% do not use final punctuation, in order to avoid confusing it with
%%% the Web address.
%%%
%%% To suppress output of a particular field, define its macro to expand
%%% to an empty string, or better, \unskip, like this:
%%%
%%% \newcommand{\showDOI}[1]{\unskip}   % LaTeX syntax
%%%
%%% \def \showDOI #1{\unskip}           % plain TeX syntax
%%%
%%% ====================================================================

\ifx \showCODEN    \undefined \def \showCODEN     #1{\unskip}     \fi
\ifx \showDOI      \undefined \def \showDOI       #1{#1}\fi
\ifx \showISBNx    \undefined \def \showISBNx     #1{\unskip}     \fi
\ifx \showISBNxiii \undefined \def \showISBNxiii  #1{\unskip}     \fi
\ifx \showISSN     \undefined \def \showISSN      #1{\unskip}     \fi
\ifx \showLCCN     \undefined \def \showLCCN      #1{\unskip}     \fi
\ifx \shownote     \undefined \def \shownote      #1{#1}          \fi
\ifx \showarticletitle \undefined \def \showarticletitle #1{#1}   \fi
\ifx \showURL      \undefined \def \showURL       {\relax}        \fi
% The following commands are used for tagged output and should be
% invisible to TeX
\providecommand\bibfield[2]{#2}
\providecommand\bibinfo[2]{#2}
\providecommand\natexlab[1]{#1}
\providecommand\showeprint[2][]{arXiv:#2}

\bibitem[\protect\citeauthoryear{Bauer}{Bauer}{2019}]%
        {bauer2019allowing}
\bibfield{author}{\bibinfo{person}{Christine Bauer}.}
  \bibinfo{year}{2019}\natexlab{}.
\newblock \showarticletitle{Allowing for equal opportunities for artists in
  music recommendation}.
\newblock \bibinfo{journal}{\emph{CoRR}}  \bibinfo{volume}{abs/1911.05395}
  (\bibinfo{year}{2019}).
\newblock
\showeprint[arXiv]{1911.05395}
\urldef\tempurl%
\url{http://arxiv.org/abs/1911.05395}
\showURL{%
\tempurl}


\bibitem[\protect\citeauthoryear{Beigi, Mosallanezhad, Guo, Alvari, Nou, and
  Liu}{Beigi et~al\mbox{.}}{2020}]%
        {beigi_privacy_aware_recommendations}
\bibfield{author}{\bibinfo{person}{Ghazaleh Beigi}, \bibinfo{person}{Ahmadreza
  Mosallanezhad}, \bibinfo{person}{Ruocheng Guo}, \bibinfo{person}{Hamidreza
  Alvari}, \bibinfo{person}{Alexander Nou}, {and} \bibinfo{person}{Huan Liu}.}
  \bibinfo{year}{2020}\natexlab{}.
\newblock \showarticletitle{Privacy-Aware Recommendation with Private-Attribute
  Protection using Adversarial Learning}. In \bibinfo{booktitle}{\emph{{WSDM}
  '20: The Thirteenth {ACM} International Conference on Web Search and Data
  Mining, Houston, TX, USA, February 3-7, 2020}},
  \bibfield{editor}{\bibinfo{person}{James Caverlee},
  \bibinfo{person}{Xia~(Ben) Hu}, \bibinfo{person}{Mounia Lalmas}, {and}
  \bibinfo{person}{Wei Wang}} (Eds.). \bibinfo{publisher}{{ACM}},
  \bibinfo{pages}{34--42}.
\newblock
\urldef\tempurl%
\url{https://doi.org/10.1145/3336191.3371832}
\showDOI{\tempurl}


\bibitem[\protect\citeauthoryear{Beigi, Shu, Guo, Wang, and Liu}{Beigi
  et~al\mbox{.}}{2019}]%
        {beigi_privacy_conserving_text}
\bibfield{author}{\bibinfo{person}{Ghazaleh Beigi}, \bibinfo{person}{Kai Shu},
  \bibinfo{person}{Ruocheng Guo}, \bibinfo{person}{Suhang Wang}, {and}
  \bibinfo{person}{Huan Liu}.} \bibinfo{year}{2019}\natexlab{}.
\newblock \showarticletitle{Privacy Preserving Text Representation Learning}.
  In \bibinfo{booktitle}{\emph{Proceedings of the 30th {ACM} Conference on
  Hypertext and Social Media, {HT} 2019, Hof, Germany, September 17-20, 2019}},
  \bibfield{editor}{\bibinfo{person}{Claus Atzenbeck}, \bibinfo{person}{Jessica
  Rubart}, {and} \bibinfo{person}{David~E. Millard}} (Eds.).
  \bibinfo{publisher}{{ACM}}, \bibinfo{pages}{275--276}.
\newblock
\urldef\tempurl%
\url{https://doi.org/10.1145/3342220.3344925}
\showDOI{\tempurl}


\bibitem[\protect\citeauthoryear{Brodersen, Ong, Stephan, and
  Buhmann}{Brodersen et~al\mbox{.}}{2010}]%
        {brodersen_balanced_accuracy}
\bibfield{author}{\bibinfo{person}{Kay~Henning Brodersen},
  \bibinfo{person}{Cheng~Soon Ong}, \bibinfo{person}{Klaas~Enno Stephan}, {and}
  \bibinfo{person}{Joachim~M. Buhmann}.} \bibinfo{year}{2010}\natexlab{}.
\newblock \showarticletitle{The Balanced Accuracy and Its Posterior
  Distribution}. In \bibinfo{booktitle}{\emph{20th International Conference on
  Pattern Recognition, {ICPR} 2010, Istanbul, Turkey, 23-26 August 2010}}.
  \bibinfo{publisher}{{IEEE} Computer Society}, \bibinfo{pages}{3121--3124}.
\newblock
\urldef\tempurl%
\url{https://doi.org/10.1109/ICPR.2010.764}
\showDOI{\tempurl}


\bibitem[\protect\citeauthoryear{Dacrema, Cremonesi, and Jannach}{Dacrema
  et~al\mbox{.}}{2019}]%
        {DBLP:conf/recsys/DacremaCJ19}
\bibfield{author}{\bibinfo{person}{Maurizio~Ferrari Dacrema},
  \bibinfo{person}{Paolo Cremonesi}, {and} \bibinfo{person}{Dietmar Jannach}.}
  \bibinfo{year}{2019}\natexlab{}.
\newblock \showarticletitle{Are we really making much progress? {A} worrying
  analysis of recent neural recommendation approaches}. In
  \bibinfo{booktitle}{\emph{Proceedings of the 13th {ACM} Conference on
  Recommender Systems, RecSys 2019, Copenhagen, Denmark, September 16-20,
  2019}}, \bibfield{editor}{\bibinfo{person}{Toine Bogers},
  \bibinfo{person}{Alan Said}, \bibinfo{person}{Peter Brusilovsky}, {and}
  \bibinfo{person}{Domonkos Tikk}} (Eds.). \bibinfo{publisher}{{ACM}},
  \bibinfo{pages}{101--109}.
\newblock
\urldef\tempurl%
\url{https://doi.org/10.1145/3298689.3347058}
\showDOI{\tempurl}


\bibitem[\protect\citeauthoryear{Deldjoo, Noia, and Merra}{Deldjoo
  et~al\mbox{.}}{2021}]%
        {DBLP:journals/csur/DeldjooNM21}
\bibfield{author}{\bibinfo{person}{Yashar Deldjoo}, \bibinfo{person}{Tommaso~Di
  Noia}, {and} \bibinfo{person}{Felice~Antonio Merra}.}
  \bibinfo{year}{2021}\natexlab{}.
\newblock \showarticletitle{A Survey on Adversarial Recommender Systems: From
  Attack/Defense Strategies to Generative Adversarial Networks}.
\newblock \bibinfo{journal}{\emph{{ACM} Comput. Surv.}} \bibinfo{volume}{54},
  \bibinfo{number}{2} (\bibinfo{year}{2021}), \bibinfo{pages}{35:1--35:38}.
\newblock
\urldef\tempurl%
\url{https://doi.org/10.1145/3439729}
\showDOI{\tempurl}


\bibitem[\protect\citeauthoryear{Ekstrand, Tian, Azpiazu, Ekstrand, Anuyah,
  McNeill, and Pera}{Ekstrand et~al\mbox{.}}{2018a}]%
        {ekstrand2018all}
\bibfield{author}{\bibinfo{person}{Michael~D. Ekstrand}, \bibinfo{person}{Mucun
  Tian}, \bibinfo{person}{Ion~Madrazo Azpiazu}, \bibinfo{person}{Jennifer~D.
  Ekstrand}, \bibinfo{person}{Oghenemaro Anuyah}, \bibinfo{person}{David
  McNeill}, {and} \bibinfo{person}{Maria~Soledad Pera}.}
  \bibinfo{year}{2018}\natexlab{a}.
\newblock \showarticletitle{All The Cool Kids, How Do They Fit In?: Popularity
  and Demographic Biases in Recommender Evaluation and Effectiveness}. In
  \bibinfo{booktitle}{\emph{Conference on Fairness, Accountability and
  Transparency, {FAT} 2018, 23-24 February 2018, New York, NY, {USA}}}
  \emph{(\bibinfo{series}{Proceedings of Machine Learning Research},
  Vol.~\bibinfo{volume}{81})}, \bibfield{editor}{\bibinfo{person}{Sorelle~A.
  Friedler} {and} \bibinfo{person}{Christo Wilson}} (Eds.).
  \bibinfo{publisher}{{PMLR}}, \bibinfo{pages}{172--186}.
\newblock
\urldef\tempurl%
\url{http://proceedings.mlr.press/v81/ekstrand18b.html}
\showURL{%
\tempurl}


\bibitem[\protect\citeauthoryear{Ekstrand, Tian, Kazi, Mehrpouyan, and
  Kluver}{Ekstrand et~al\mbox{.}}{2018b}]%
        {ekstrand2018exploring}
\bibfield{author}{\bibinfo{person}{Michael~D. Ekstrand}, \bibinfo{person}{Mucun
  Tian}, \bibinfo{person}{Mohammed R.~Imran Kazi}, \bibinfo{person}{Hoda
  Mehrpouyan}, {and} \bibinfo{person}{Daniel Kluver}.}
  \bibinfo{year}{2018}\natexlab{b}.
\newblock \showarticletitle{Exploring author gender in book rating and
  recommendation}. In \bibinfo{booktitle}{\emph{Proceedings of the 12th {ACM}
  Conference on Recommender Systems, RecSys 2018, Vancouver, BC, Canada,
  October 2-7, 2018}}, \bibfield{editor}{\bibinfo{person}{Sole Pera},
  \bibinfo{person}{Michael~D. Ekstrand}, \bibinfo{person}{Xavier Amatriain},
  {and} \bibinfo{person}{John O'Donovan}} (Eds.). \bibinfo{publisher}{{ACM}},
  \bibinfo{pages}{242--250}.
\newblock
\urldef\tempurl%
\url{https://doi.org/10.1145/3240323.3240373}
\showDOI{\tempurl}


\bibitem[\protect\citeauthoryear{Elahi, Jannach, Skj{\ae}rven, Knudsen,
  Sj{\o}vaag, Tolonen, Holmstad, Pipkin, Throndsen, Stenbom, Fiskerud, Oesch,
  Vredenberg, and Trattner}{Elahi et~al\mbox{.}}{2021}]%
        {elahi2021towards}
\bibfield{author}{\bibinfo{person}{Mehdi Elahi}, \bibinfo{person}{Dietmar
  Jannach}, \bibinfo{person}{Lars Skj{\ae}rven}, \bibinfo{person}{Erik
  Knudsen}, \bibinfo{person}{Helle Sj{\o}vaag}, \bibinfo{person}{Kristian
  Tolonen}, \bibinfo{person}{{\O}yvind Holmstad}, \bibinfo{person}{Igor
  Pipkin}, \bibinfo{person}{Eivind Throndsen}, \bibinfo{person}{Agnes Stenbom},
  \bibinfo{person}{Eivind Fiskerud}, \bibinfo{person}{Adrian Oesch},
  \bibinfo{person}{Loek Vredenberg}, {and} \bibinfo{person}{Christoph
  Trattner}.} \bibinfo{year}{2021}\natexlab{}.
\newblock \showarticletitle{Towards responsible media recommendation}.
\newblock \bibinfo{journal}{\emph{AI and Ethics}} (\bibinfo{date}{02 Nov}
  \bibinfo{year}{2021}).
\newblock
\showISSN{2730-5961}
\urldef\tempurl%
\url{https://doi.org/10.1007/s43681-021-00107-7}
\showDOI{\tempurl}


\bibitem[\protect\citeauthoryear{Elazar and Goldberg}{Elazar and
  Goldberg}{2018}]%
        {elazar-goldberg-2018-adversarial}
\bibfield{author}{\bibinfo{person}{Yanai Elazar} {and} \bibinfo{person}{Yoav
  Goldberg}.} \bibinfo{year}{2018}\natexlab{}.
\newblock \showarticletitle{Adversarial Removal of Demographic Attributes from
  Text Data}. In \bibinfo{booktitle}{\emph{Proceedings of the 2018 Conference
  on Empirical Methods in Natural Language Processing, Brussels, Belgium,
  October 31 - November 4, 2018}}, \bibfield{editor}{\bibinfo{person}{Ellen
  Riloff}, \bibinfo{person}{David Chiang}, \bibinfo{person}{Julia Hockenmaier},
  {and} \bibinfo{person}{Jun'ichi Tsujii}} (Eds.).
  \bibinfo{publisher}{Association for Computational Linguistics},
  \bibinfo{pages}{11--21}.
\newblock
\urldef\tempurl%
\url{https://doi.org/10.18653/v1/d18-1002}
\showDOI{\tempurl}


\bibitem[\protect\citeauthoryear{Esp{\'{\i}}n{-}Noboa, Wagner, Strohmaier, and
  Karimi}{Esp{\'{\i}}n{-}Noboa et~al\mbox{.}}{2021}]%
        {espin2022inequality}
\bibfield{author}{\bibinfo{person}{Lisette Esp{\'{\i}}n{-}Noboa},
  \bibinfo{person}{Claudia Wagner}, \bibinfo{person}{Markus Strohmaier}, {and}
  \bibinfo{person}{Fariba Karimi}.} \bibinfo{year}{2021}\natexlab{}.
\newblock \showarticletitle{Inequality and Inequity in Network-based Ranking
  and Recommendation Algorithms}.
\newblock \bibinfo{journal}{\emph{CoRR}}  \bibinfo{volume}{abs/2110.00072}
  (\bibinfo{year}{2021}).
\newblock
\showeprint[arXiv]{2110.00072}
\urldef\tempurl%
\url{https://arxiv.org/abs/2110.00072}
\showURL{%
\tempurl}


\bibitem[\protect\citeauthoryear{Ganin and Lempitsky}{Ganin and
  Lempitsky}{2015}]%
        {ganing_gradient_reversal}
\bibfield{author}{\bibinfo{person}{Yaroslav Ganin} {and}
  \bibinfo{person}{Victor~S. Lempitsky}.} \bibinfo{year}{2015}\natexlab{}.
\newblock \showarticletitle{Unsupervised Domain Adaptation by Backpropagation}.
  In \bibinfo{booktitle}{\emph{Proceedings of the 32nd International Conference
  on Machine Learning, {ICML} 2015, Lille, France, 6-11 July 2015}}
  \emph{(\bibinfo{series}{{JMLR} Workshop and Conference Proceedings},
  Vol.~\bibinfo{volume}{37})}, \bibfield{editor}{\bibinfo{person}{Francis~R.
  Bach} {and} \bibinfo{person}{David~M. Blei}} (Eds.).
  \bibinfo{publisher}{JMLR.org}, \bibinfo{pages}{1180--1189}.
\newblock
\urldef\tempurl%
\url{http://proceedings.mlr.press/v37/ganin15.html}
\showURL{%
\tempurl}


\bibitem[\protect\citeauthoryear{Goodfellow, Pouget{-}Abadie, Mirza, Xu,
  Warde{-}Farley, Ozair, Courville, and Bengio}{Goodfellow
  et~al\mbox{.}}{2014}]%
        {NIPS2014_5ca3e9b1}
\bibfield{author}{\bibinfo{person}{Ian~J. Goodfellow}, \bibinfo{person}{Jean
  Pouget{-}Abadie}, \bibinfo{person}{Mehdi Mirza}, \bibinfo{person}{Bing Xu},
  \bibinfo{person}{David Warde{-}Farley}, \bibinfo{person}{Sherjil Ozair},
  \bibinfo{person}{Aaron~C. Courville}, {and} \bibinfo{person}{Yoshua Bengio}.}
  \bibinfo{year}{2014}\natexlab{}.
\newblock \showarticletitle{Generative Adversarial Nets}. In
  \bibinfo{booktitle}{\emph{Advances in Neural Information Processing Systems
  27: Annual Conference on Neural Information Processing Systems 2014, December
  8-13 2014, Montreal, Quebec, Canada}},
  \bibfield{editor}{\bibinfo{person}{Zoubin Ghahramani}, \bibinfo{person}{Max
  Welling}, \bibinfo{person}{Corinna Cortes}, \bibinfo{person}{Neil~D.
  Lawrence}, {and} \bibinfo{person}{Kilian~Q. Weinberger}} (Eds.).
  \bibinfo{pages}{2672--2680}.
\newblock
\urldef\tempurl%
\url{https://proceedings.neurips.cc/paper/2014/hash/5ca3e9b122f61f8f06494c97b1afccf3-Abstract.html}
\showURL{%
\tempurl}


\bibitem[\protect\citeauthoryear{Harper and Konstan}{Harper and
  Konstan}{2015}]%
        {harper_movielens_2015}
\bibfield{author}{\bibinfo{person}{F.~Maxwell Harper} {and}
  \bibinfo{person}{Joseph~A. Konstan}.} \bibinfo{year}{2015}\natexlab{}.
\newblock \showarticletitle{The MovieLens Datasets: History and Context}.
\newblock \bibinfo{journal}{\emph{ACM Trans. Interact. Intell. Syst.}}
  \bibinfo{volume}{5}, \bibinfo{number}{4}, Article \bibinfo{articleno}{19}
  (\bibinfo{date}{dec} \bibinfo{year}{2015}), \bibinfo{numpages}{19}~pages.
\newblock
\showISSN{2160-6455}
\urldef\tempurl%
\url{https://doi.org/10.1145/2827872}
\showDOI{\tempurl}


\bibitem[\protect\citeauthoryear{Kingma and Welling}{Kingma and
  Welling}{2014}]%
        {journals/corr/KingmaW13}
\bibfield{author}{\bibinfo{person}{Diederik~P. Kingma} {and}
  \bibinfo{person}{Max Welling}.} \bibinfo{year}{2014}\natexlab{}.
\newblock \showarticletitle{Auto-Encoding Variational Bayes}. In
  \bibinfo{booktitle}{\emph{2nd International Conference on Learning
  Representations, {ICLR} 2014, Banff, AB, Canada, April 14-16, 2014,
  Conference Track Proceedings}}, \bibfield{editor}{\bibinfo{person}{Yoshua
  Bengio} {and} \bibinfo{person}{Yann LeCun}} (Eds.).
\newblock
\urldef\tempurl%
\url{http://arxiv.org/abs/1312.6114}
\showURL{%
\tempurl}


\bibitem[\protect\citeauthoryear{Krieg, Parada-Cabaleiro, Medicus, Lesota,
  Schedl, and Rekabsaz}{Krieg et~al\mbox{.}}{2022a}]%
        {krieg2022grep}
\bibfield{author}{\bibinfo{person}{Klara Krieg}, \bibinfo{person}{Emilia
  Parada-Cabaleiro}, \bibinfo{person}{Gertraud Medicus}, \bibinfo{person}{Oleg
  Lesota}, \bibinfo{person}{Markus Schedl}, {and} \bibinfo{person}{Navid
  Rekabsaz}.} \bibinfo{year}{2022}\natexlab{a}.
\newblock \showarticletitle{{Grep-BiasIR}: A Dataset for Investigating Gender
  Representation-Bias in Information Retrieval Results}.
\newblock \bibinfo{journal}{\emph{arXiv preprint arXiv:2201.07754}}
  (\bibinfo{year}{2022}).
\newblock


\bibitem[\protect\citeauthoryear{Krieg, Parada-Cabaleiro, Schedl, and
  Rekabsaz}{Krieg et~al\mbox{.}}{2022b}]%
        {krieg2022do}
\bibfield{author}{\bibinfo{person}{Klara Krieg}, \bibinfo{person}{Emilia
  Parada-Cabaleiro}, \bibinfo{person}{Markus Schedl}, {and}
  \bibinfo{person}{Navid Rekabsaz}.} \bibinfo{year}{2022}\natexlab{b}.
\newblock \showarticletitle{Do Perceived Gender Biases in Retrieval Results
  Affect Relevance Judgements?}. In \bibinfo{booktitle}{\emph{Proceedings of
  the Workshop on Algorithmic Bias in Search and Recommendation at the European
  Conference on Information Retrieval ({ECIR-BIAS} 2022)}}.
\newblock


\bibitem[\protect\citeauthoryear{Li, Chen, Xu, Ge, and Zhang}{Li
  et~al\mbox{.}}{2021}]%
        {li_PersonalizedFairnessBased_2021}
\bibfield{author}{\bibinfo{person}{Yunqi Li}, \bibinfo{person}{Hanxiong Chen},
  \bibinfo{person}{Shuyuan Xu}, \bibinfo{person}{Yingqiang Ge}, {and}
  \bibinfo{person}{Yongfeng Zhang}.} \bibinfo{year}{2021}\natexlab{}.
\newblock \showarticletitle{Towards {{Personalized Fairness}} Based on {{Causal
  Notion}}}. In \bibinfo{booktitle}{\emph{Proceedings of the 44th
  {{International ACM SIGIR Conference}} on {{Research}} and {{Development}} in
  {{Information Retrieval}}}}. \bibinfo{publisher}{{ACM}},
  \bibinfo{address}{{Virtual Event Canada}}, \bibinfo{pages}{1054--1063}.
\newblock
\showISBNx{978-1-4503-8037-9}
\urldef\tempurl%
\url{https://doi.org/10.1145/3404835.3462966}
\showDOI{\tempurl}


\bibitem[\protect\citeauthoryear{Liang, Krishnan, Hoffman, and Jebara}{Liang
  et~al\mbox{.}}{2018}]%
        {liang_mult_vae}
\bibfield{author}{\bibinfo{person}{Dawen Liang}, \bibinfo{person}{Rahul~G.
  Krishnan}, \bibinfo{person}{Matthew~D. Hoffman}, {and} \bibinfo{person}{Tony
  Jebara}.} \bibinfo{year}{2018}\natexlab{}.
\newblock \showarticletitle{Variational Autoencoders for Collaborative
  Filtering}. In \bibinfo{booktitle}{\emph{Proceedings of the 2018 World Wide
  Web Conference on World Wide Web, {WWW} 2018, Lyon, France, April 23-27,
  2018}}, \bibfield{editor}{\bibinfo{person}{Pierre{-}Antoine Champin},
  \bibinfo{person}{Fabien Gandon}, \bibinfo{person}{Mounia Lalmas}, {and}
  \bibinfo{person}{Panagiotis~G. Ipeirotis}} (Eds.).
  \bibinfo{publisher}{{ACM}}, \bibinfo{pages}{689--698}.
\newblock
\urldef\tempurl%
\url{https://doi.org/10.1145/3178876.3186150}
\showDOI{\tempurl}


\bibitem[\protect\citeauthoryear{McNemar}{McNemar}{1947}]%
        {McNemar1947}
\bibfield{author}{\bibinfo{person}{Quinn McNemar}.}
  \bibinfo{year}{1947}\natexlab{}.
\newblock \showarticletitle{Note on the sampling error of the difference
  between correlated proportions or percentages}.
\newblock \bibinfo{journal}{\emph{Psychometrika}} \bibinfo{volume}{12},
  \bibinfo{number}{2} (\bibinfo{date}{01 Jun} \bibinfo{year}{1947}),
  \bibinfo{pages}{153--157}.
\newblock
\showISSN{1860-0980}
\urldef\tempurl%
\url{https://doi.org/10.1007/BF02295996}
\showDOI{\tempurl}


\bibitem[\protect\citeauthoryear{Melchiorre, Rekabsaz, Parada{-}Cabaleiro,
  Brandl, Lesota, and Schedl}{Melchiorre et~al\mbox{.}}{2021}]%
        {melchiorre_gender_fairness}
\bibfield{author}{\bibinfo{person}{Alessandro~B. Melchiorre},
  \bibinfo{person}{Navid Rekabsaz}, \bibinfo{person}{Emilia
  Parada{-}Cabaleiro}, \bibinfo{person}{Stefan Brandl}, \bibinfo{person}{Oleg
  Lesota}, {and} \bibinfo{person}{Markus Schedl}.}
  \bibinfo{year}{2021}\natexlab{}.
\newblock \showarticletitle{Investigating gender fairness of recommendation
  algorithms in the music domain}.
\newblock \bibinfo{journal}{\emph{Inf. Process. Manag.}} \bibinfo{volume}{58},
  \bibinfo{number}{5} (\bibinfo{year}{2021}), \bibinfo{pages}{102666}.
\newblock
\urldef\tempurl%
\url{https://doi.org/10.1016/j.ipm.2021.102666}
\showDOI{\tempurl}


\bibitem[\protect\citeauthoryear{Meng, McCreadie, Macdonald, and Ounis}{Meng
  et~al\mbox{.}}{2020}]%
        {meng_data_splitting_strategies}
\bibfield{author}{\bibinfo{person}{Zaiqiao Meng}, \bibinfo{person}{Richard
  McCreadie}, \bibinfo{person}{Craig Macdonald}, {and} \bibinfo{person}{Iadh
  Ounis}.} \bibinfo{year}{2020}\natexlab{}.
\newblock \showarticletitle{Exploring Data Splitting Strategies for the
  Evaluation of Recommendation Models}. In \bibinfo{booktitle}{\emph{RecSys
  2020: Fourteenth {ACM} Conference on Recommender Systems, Virtual Event,
  Brazil, September 22-26, 2020}}, \bibfield{editor}{\bibinfo{person}{Rodrygo
  L.~T. Santos}, \bibinfo{person}{Leandro~Balby Marinho},
  \bibinfo{person}{Elizabeth~M. Daly}, \bibinfo{person}{Li~Chen},
  \bibinfo{person}{Kim Falk}, \bibinfo{person}{Noam Koenigstein}, {and}
  \bibinfo{person}{Edleno~Silva de~Moura}} (Eds.). \bibinfo{publisher}{{ACM}},
  \bibinfo{pages}{681--686}.
\newblock
\urldef\tempurl%
\url{https://doi.org/10.1145/3383313.3418479}
\showDOI{\tempurl}


\bibitem[\protect\citeauthoryear{Rekabsaz, Kopeinik, and Schedl}{Rekabsaz
  et~al\mbox{.}}{2021}]%
        {rekabsaz_societal_biases_adv_bert}
\bibfield{author}{\bibinfo{person}{Navid Rekabsaz}, \bibinfo{person}{Simone
  Kopeinik}, {and} \bibinfo{person}{Markus Schedl}.}
  \bibinfo{year}{2021}\natexlab{}.
\newblock \showarticletitle{Societal Biases in Retrieved Contents: Measurement
  Framework and Adversarial Mitigation of {BERT} Rankers}. In
  \bibinfo{booktitle}{\emph{{SIGIR} '21: The 44th International {ACM} {SIGIR}
  Conference on Research and Development in Information Retrieval, Virtual
  Event, Canada, July 11-15, 2021}},
  \bibfield{editor}{\bibinfo{person}{Fernando Diaz}, \bibinfo{person}{Chirag
  Shah}, \bibinfo{person}{Torsten Suel}, \bibinfo{person}{Pablo Castells},
  \bibinfo{person}{Rosie Jones}, {and} \bibinfo{person}{Tetsuya Sakai}} (Eds.).
  \bibinfo{publisher}{{ACM}}, \bibinfo{pages}{306--316}.
\newblock
\urldef\tempurl%
\url{https://doi.org/10.1145/3404835.3462949}
\showDOI{\tempurl}


\bibitem[\protect\citeauthoryear{Rekabsaz and Schedl}{Rekabsaz and
  Schedl}{2020}]%
        {rekabsaz2020do}
\bibfield{author}{\bibinfo{person}{Navid Rekabsaz} {and}
  \bibinfo{person}{Markus Schedl}.} \bibinfo{year}{2020}\natexlab{}.
\newblock \showarticletitle{Do Neural Ranking Models Intensify Gender Bias?}.
  In \bibinfo{booktitle}{\emph{Proceedings of the 43rd International {ACM}
  {SIGIR} conference on research and development in Information Retrieval,
  {SIGIR} 2020, Virtual Event, China, July 25-30, 2020}},
  \bibfield{editor}{\bibinfo{person}{Jimmy Huang}, \bibinfo{person}{Yi~Chang},
  \bibinfo{person}{Xueqi Cheng}, \bibinfo{person}{Jaap Kamps},
  \bibinfo{person}{Vanessa Murdock}, \bibinfo{person}{Ji{-}Rong Wen}, {and}
  \bibinfo{person}{Yiqun Liu}} (Eds.). \bibinfo{publisher}{{ACM}},
  \bibinfo{pages}{2065--2068}.
\newblock
\urldef\tempurl%
\url{https://doi.org/10.1145/3397271.3401280}
\showDOI{\tempurl}


\bibitem[\protect\citeauthoryear{Rey and Neuh{\"{a}}user}{Rey and
  Neuh{\"{a}}user}{2011}]%
        {DBLP:reference/stat/ReyN11}
\bibfield{author}{\bibinfo{person}{Denise Rey} {and} \bibinfo{person}{Markus
  Neuh{\"{a}}user}.} \bibinfo{year}{2011}\natexlab{}.
\newblock \showarticletitle{Wilcoxon-Signed-Rank Test}.
\newblock In \bibinfo{booktitle}{\emph{International Encyclopedia of
  Statistical Science}}, \bibfield{editor}{\bibinfo{person}{Miodrag Lovric}}
  (Ed.). \bibinfo{publisher}{Springer}, \bibinfo{pages}{1658--1659}.
\newblock
\urldef\tempurl%
\url{https://doi.org/10.1007/978-3-642-04898-2\_616}
\showDOI{\tempurl}


\bibitem[\protect\citeauthoryear{Schedl, Brandl, Lesota, Parada-Cabaleiro,
  Penz, and Rekabsaz}{Schedl et~al\mbox{.}}{2022}]%
        {schedl2022lfm}
\bibfield{author}{\bibinfo{person}{Markus Schedl}, \bibinfo{person}{Stefan
  Brandl}, \bibinfo{person}{Oleg Lesota}, \bibinfo{person}{Emilia
  Parada-Cabaleiro}, \bibinfo{person}{David Penz}, {and} \bibinfo{person}{Navid
  Rekabsaz}.} \bibinfo{year}{2022}\natexlab{}.
\newblock \showarticletitle{LFM-2b: A Dataset of Enriched Music Listening
  Events for Recommender Systems Research and Fairness Analysis}. In
  \bibinfo{booktitle}{\emph{ACM SIGIR Conference on Human Information
  Interaction and Retrieval}}. \bibinfo{pages}{337--341}.
\newblock


\bibitem[\protect\citeauthoryear{Schedl, Hauger, Farrahi, and Tkalcic}{Schedl
  et~al\mbox{.}}{2015}]%
        {schedl2015influence}
\bibfield{author}{\bibinfo{person}{Markus Schedl}, \bibinfo{person}{David
  Hauger}, \bibinfo{person}{Katayoun Farrahi}, {and} \bibinfo{person}{Marko
  Tkalcic}.} \bibinfo{year}{2015}\natexlab{}.
\newblock \showarticletitle{On the Influence of User Characteristics on Music
  Recommendation Algorithms}. In \bibinfo{booktitle}{\emph{Advances in
  Information Retrieval - 37th European Conference on {IR} Research, {ECIR}
  2015, Vienna, Austria, March 29 - April 2, 2015. Proceedings}}
  \emph{(\bibinfo{series}{Lecture Notes in Computer Science},
  Vol.~\bibinfo{volume}{9022})}, \bibfield{editor}{\bibinfo{person}{Allan
  Hanbury}, \bibinfo{person}{Gabriella Kazai}, \bibinfo{person}{Andreas
  Rauber}, {and} \bibinfo{person}{Norbert Fuhr}} (Eds.).
  \bibinfo{pages}{339--345}.
\newblock
\urldef\tempurl%
\url{https://doi.org/10.1007/978-3-319-16354-3\_37}
\showDOI{\tempurl}


\bibitem[\protect\citeauthoryear{Wang and Chen}{Wang and Chen}{2021}]%
        {10.1145/3460231.3474244}
\bibfield{author}{\bibinfo{person}{Ningxia Wang} {and} \bibinfo{person}{Li
  Chen}.} \bibinfo{year}{2021}\natexlab{}.
\newblock \showarticletitle{User Bias in Beyond-Accuracy Measurement of
  Recommendation Algorithms}. In \bibinfo{booktitle}{\emph{RecSys '21:
  Fifteenth {ACM} Conference on Recommender Systems, Amsterdam, The
  Netherlands, 27 September 2021 - 1 October 2021}},
  \bibfield{editor}{\bibinfo{person}{Humberto Jes{\'{u}}s~Corona
  Pamp{\'{\i}}n}, \bibinfo{person}{Martha~A. Larson},
  \bibinfo{person}{Martijn~C. Willemsen}, \bibinfo{person}{Joseph~A. Konstan},
  \bibinfo{person}{Julian~J. McAuley}, \bibinfo{person}{Jean
  Garcia{-}Gathright}, \bibinfo{person}{Bouke Huurnink}, {and}
  \bibinfo{person}{Even Oldridge}} (Eds.). \bibinfo{publisher}{{ACM}},
  \bibinfo{pages}{133--142}.
\newblock
\urldef\tempurl%
\url{https://doi.org/10.1145/3460231.3474244}
\showDOI{\tempurl}


\bibitem[\protect\citeauthoryear{Weinsberg, Bhagat, Ioannidis, and
  Taft}{Weinsberg et~al\mbox{.}}{2012}]%
        {weinsberg_blurme_sensitive_information_in_history}
\bibfield{author}{\bibinfo{person}{Udi Weinsberg}, \bibinfo{person}{Smriti
  Bhagat}, \bibinfo{person}{Stratis Ioannidis}, {and} \bibinfo{person}{Nina
  Taft}.} \bibinfo{year}{2012}\natexlab{}.
\newblock \showarticletitle{BlurMe: inferring and obfuscating user gender based
  on ratings}. In \bibinfo{booktitle}{\emph{Sixth {ACM} Conference on
  Recommender Systems, RecSys '12, Dublin, Ireland, September 9-13, 2012}},
  \bibfield{editor}{\bibinfo{person}{Padraig Cunningham},
  \bibinfo{person}{Neil~J. Hurley}, \bibinfo{person}{Ido Guy}, {and}
  \bibinfo{person}{Sarabjot~Singh Anand}} (Eds.). \bibinfo{publisher}{{ACM}},
  \bibinfo{pages}{195--202}.
\newblock
\urldef\tempurl%
\url{https://doi.org/10.1145/2365952.2365989}
\showDOI{\tempurl}


\bibitem[\protect\citeauthoryear{Wu, Wu, Wang, Huang, and Xie}{Wu
  et~al\mbox{.}}{2021b}]%
        {DBLP:conf/aaai/WuWWH021}
\bibfield{author}{\bibinfo{person}{Chuhan Wu}, \bibinfo{person}{Fangzhao Wu},
  \bibinfo{person}{Xiting Wang}, \bibinfo{person}{Yongfeng Huang}, {and}
  \bibinfo{person}{Xing Xie}.} \bibinfo{year}{2021}\natexlab{b}.
\newblock \showarticletitle{Fairness-aware News Recommendation with Decomposed
  Adversarial Learning}. In \bibinfo{booktitle}{\emph{Thirty-Fifth {AAAI}
  Conference on Artificial Intelligence, {AAAI} 2021, Thirty-Third Conference
  on Innovative Applications of Artificial Intelligence, {IAAI} 2021, The
  Eleventh Symposium on Educational Advances in Artificial Intelligence, {EAAI}
  2021, Virtual Event, February 2-9, 2021}}. \bibinfo{publisher}{{AAAI} Press},
  \bibinfo{pages}{4462--4469}.
\newblock
\urldef\tempurl%
\url{https://ojs.aaai.org/index.php/AAAI/article/view/16573}
\showURL{%
\tempurl}


\bibitem[\protect\citeauthoryear{Wu, Chen, Shao, Hong, Wang, and Wang}{Wu
  et~al\mbox{.}}{2021a}]%
        {wu_LearningFairRepresentations_2021}
\bibfield{author}{\bibinfo{person}{Le Wu}, \bibinfo{person}{Lei Chen},
  \bibinfo{person}{Pengyang Shao}, \bibinfo{person}{Richang Hong},
  \bibinfo{person}{Xiting Wang}, {and} \bibinfo{person}{Meng Wang}.}
  \bibinfo{year}{2021}\natexlab{a}.
\newblock \showarticletitle{Learning {{Fair Representations}} for
  {{Recommendation}}: {{A Graph-based Perspective}}}. In
  \bibinfo{booktitle}{\emph{Proceedings of the {{Web Conference}} 2021}}.
  \bibinfo{publisher}{{ACM}}, \bibinfo{address}{{Ljubljana Slovenia}},
  \bibinfo{pages}{2198--2208}.
\newblock
\showISBNx{978-1-4503-8312-7}
\urldef\tempurl%
\url{https://doi.org/10.1145/3442381.3450015}
\showDOI{\tempurl}


\bibitem[\protect\citeauthoryear{Xie, Dai, Du, Hovy, and Neubig}{Xie
  et~al\mbox{.}}{2017}]%
        {xie2018controllable}
\bibfield{author}{\bibinfo{person}{Qizhe Xie}, \bibinfo{person}{Zihang Dai},
  \bibinfo{person}{Yulun Du}, \bibinfo{person}{Eduard~H. Hovy}, {and}
  \bibinfo{person}{Graham Neubig}.} \bibinfo{year}{2017}\natexlab{}.
\newblock \showarticletitle{Controllable Invariance through Adversarial Feature
  Learning}. In \bibinfo{booktitle}{\emph{Advances in Neural Information
  Processing Systems 30: Annual Conference on Neural Information Processing
  Systems 2017, December 4-9, 2017, Long Beach, CA, {USA}}},
  \bibfield{editor}{\bibinfo{person}{Isabelle Guyon}, \bibinfo{person}{Ulrike
  von Luxburg}, \bibinfo{person}{Samy Bengio}, \bibinfo{person}{Hanna~M.
  Wallach}, \bibinfo{person}{Rob Fergus}, \bibinfo{person}{S.~V.~N.
  Vishwanathan}, {and} \bibinfo{person}{Roman Garnett}} (Eds.).
  \bibinfo{pages}{585--596}.
\newblock
\urldef\tempurl%
\url{https://proceedings.neurips.cc/paper/2017/hash/8cb22bdd0b7ba1ab13d742e22eed8da2-Abstract.html}
\showURL{%
\tempurl}


\bibitem[\protect\citeauthoryear{Zerveas, Rekabsaz, Cohen, and
  Eickhoff}{Zerveas et~al\mbox{.}}{2022}]%
        {zerveas2022mitigating}
\bibfield{author}{\bibinfo{person}{George Zerveas}, \bibinfo{person}{Navid
  Rekabsaz}, \bibinfo{person}{Daniel Cohen}, {and} \bibinfo{person}{Carsten
  Eickhoff}.} \bibinfo{year}{2022}\natexlab{}.
\newblock \showarticletitle{Mitigating bias in search results through set-based
  document reranking and neutrality regularization}. In
  \bibinfo{booktitle}{\emph{Proceedings of the 45th International {ACM} {SIGIR}
  conference on research and development in Information Retrieval, {SIGIR}
  2022}}. \bibinfo{publisher}{{ACM}}.
\newblock


\bibitem[\protect\citeauthoryear{Zhang, Lemoine, and Mitchell}{Zhang
  et~al\mbox{.}}{2018}]%
        {zhang_MitigatingUnwantedBiases_2018}
\bibfield{author}{\bibinfo{person}{Brian~Hu Zhang}, \bibinfo{person}{Blake
  Lemoine}, {and} \bibinfo{person}{Margaret Mitchell}.}
  \bibinfo{year}{2018}\natexlab{}.
\newblock \showarticletitle{Mitigating {{Unwanted Biases}} with {{Adversarial
  Learning}}}. In \bibinfo{booktitle}{\emph{Proceedings of the 2018
  {{AAAI}}/{{ACM Conference}} on {{AI}}, {{Ethics}}, and {{Society}}}}.
  \bibinfo{publisher}{{ACM}}, \bibinfo{address}{{New Orleans LA USA}},
  \bibinfo{pages}{335--340}.
\newblock
\showISBNx{978-1-4503-6012-8}
\urldef\tempurl%
\url{https://doi.org/10.1145/3278721.3278779}
\showDOI{\tempurl}


\bibitem[\protect\citeauthoryear{Zhu, Wang, and Caverlee}{Zhu
  et~al\mbox{.}}{2020}]%
        {DBLP:conf/sigir/ZhuWC20}
\bibfield{author}{\bibinfo{person}{Ziwei Zhu}, \bibinfo{person}{Jianling Wang},
  {and} \bibinfo{person}{James Caverlee}.} \bibinfo{year}{2020}\natexlab{}.
\newblock \showarticletitle{Measuring and Mitigating Item Under-Recommendation
  Bias in Personalized Ranking Systems}. In
  \bibinfo{booktitle}{\emph{Proceedings of the 43rd International {ACM} {SIGIR}
  conference on research and development in Information Retrieval, {SIGIR}
  2020, Virtual Event, China, July 25-30, 2020}},
  \bibfield{editor}{\bibinfo{person}{Jimmy Huang}, \bibinfo{person}{Yi~Chang},
  \bibinfo{person}{Xueqi Cheng}, \bibinfo{person}{Jaap Kamps},
  \bibinfo{person}{Vanessa Murdock}, \bibinfo{person}{Ji{-}Rong Wen}, {and}
  \bibinfo{person}{Yiqun Liu}} (Eds.). \bibinfo{publisher}{{ACM}},
  \bibinfo{pages}{449--458}.
\newblock
\urldef\tempurl%
\url{https://doi.org/10.1145/3397271.3401177}
\showDOI{\tempurl}


\end{thebibliography}

%%
%% If your work has an appendix, this is the place to put it.

\end{document}